\documentclass[12pt]{article}
\usepackage[latin1]{inputenc}

\usepackage{amsmath}
\usepackage{color}
\usepackage{amsfonts}
\usepackage{amssymb}
\usepackage{graphicx}
\usepackage{geometry}
\usepackage{amssymb,epsfig,subfigure}
\usepackage{hyperref}
\usepackage{comment}
\usepackage{tabularx}
\usepackage{bm}
\usepackage{euscript}
\usepackage{graphicx}
\usepackage{color}
\usepackage{amsfonts}
\usepackage{exscale}
\usepackage{amsbsy}
\usepackage{subfigure}
\usepackage{textcomp}
\usepackage{comment}
\usepackage{hyperref}
\usepackage{slashed}
\usepackage{authblk}
\usepackage{mathtools}
\usepackage{tabularx}
\usepackage{bm}
\usepackage{euscript}
\usepackage{graphicx}
\usepackage{color}
\usepackage{amsfonts}
\usepackage{exscale}
\usepackage{amsbsy}
\usepackage{subfigure}
\usepackage{textcomp}
\usepackage{comment}
\usepackage{hyperref}

\def\be{\begin{equation}}
\def\ee{\end{equation}}
\def\bea{\begin{eqnarray}}
\def\eea{\end{eqnarray}}
\def\ba#1\ea{\begin{align}#1\end{align}}
\def\bg#1\eg{\begin{gather}#1\end{gather}}
\def\bm#1\em{\begin{multline}#1\end{multline}}
\def\bmd#1\emd{\begin{multlined}#1\end{multlined}}

\usepackage[numbers,sort&compress]{natbib}

\def\simge{
    \mathrel{\rlap{\raise 0.511ex 
        \hbox{$>$}}{\lower 0.511ex \hbox{$\sim$}}}}

\def\simle{
    \mathrel{\rlap{\raise 0.511ex 
        \hbox{$<$}}{\lower 0.511ex \hbox{$\sim$}}}}

\makeatletter
\renewcommand\section{\@startsection {section}{1}{\z@}%
                                 {-3.5ex \@plus -1ex \@minus -.2ex}
                                   {2.3ex \@plus.2ex}%
                                   {\normalfont\large\bfseries}}
\renewcommand\subsection{\@startsection{subsection}{2}{\z@}%
                                   {-3.25ex\@plus -1ex \@minus -.2ex}%
                                     {1.5ex \@plus .2ex}%
                                     {\normalfont\bfseries}}
\renewcommand\subsubsection{\@startsection{subsubsection}{3}{\z@}%
                                   {-3.25ex\@plus -1ex \@minus -.2ex}%
                                     {1.5ex \@plus .2ex}%
                                     {\normalfont\itshape}}
\makeatother

\def\pplogo{\vbox{\kern-\headheight\kern -29pt
\halign{##&##\hfil\cr&{\ppnumber}\cr\rule{0pt}{2.5ex}&\ppdate\cr}}}
\makeatletter
\def\ps@firstpage{\ps@empty \def\@oddhead{\hss\pplogo}%
  \let\@evenhead\@oddhead 
}
\def\maketitle{\par
 \begingroup
 \def\thefootnote{\fnsymbol{footnote}}
 \def\@makefnmark{\hbox{$^{\@thefnmark}$\hss}}
 \if@twocolumn
 \twocolumn[\@maketitle]
 \else \newpage
 \global\@topnum\z@ \@maketitle \fi\thispagestyle{firstpage}\@thanks
 \endgroup
 \setcounter{footnote}{0}
 \let\maketitle\relax
 \let\@maketitle\relax
 \gdef\@thanks{}\gdef\@author{}\gdef\@title{}\let\thanks\relax}
\makeatother

\numberwithin{equation}{section}

\begin{document}

\setcounter{page}0
\def\ppnumber{\vbox{\baselineskip14pt
}}

\def\ppdate{
} \date{\today}

\title{\bf Fluctuations and magnetoresistance oscillations near the half-filled Landau level
\vskip 0.5cm}
\author{Amartya Mitra}
\author{Michael Mulligan}
\affil{\small \it Department of Physics and Astronomy, University of California,
Riverside, CA 92511, USA}

\bigskip

\maketitle

\begin{abstract}
We study theoretically the magnetoresistance oscillations near a half-filled lowest Landau level ($\nu = 1/2$) that result from the presence of a periodic one-dimensional electrostatic potential.
We use the Dirac composite fermion theory of Son [\href{http://dx.doi.org/10.1103/PhysRevX.5.031027} {Phys. Rev. X 5 031027 (2015)}], where the $\nu=1/2$ state is described by a $(2+1)$-dimensional theory of quantum electrodynamics.
We extend previous work that studied these oscillations in the mean-field limit by considering the effects of gauge field fluctuations within a large flavor approximation.
A self-consistent analysis of the resulting Schwinger--Dyson equations suggests that fluctuations dynamically generate a Chern-Simons term for the gauge field and a magnetic field-dependent mass for the Dirac composite fermions away from $\nu=1/2$.
We show how this mass results in a shift of the locations of the oscillation minima that improves the comparison with experiment [Kamburov et.~al., \href{https://journals.aps.org/prl/abstract/10.1103/PhysRevLett.113.196801} {Phys.~Rev.~Lett.~113, 196801 (2014)}].
The temperature-dependent amplitude of these oscillations may enable an alternative way to measure this mass.
This amplitude may also help distinguish the Dirac and Halperin, Lee, and Read composite fermion theories of the half-filled Landau level.
\end{abstract}
\bigskip

\newpage

\tableofcontents

\newpage

\vskip 1cm

\section{Introduction and summary}

\subsection{Motivation}

In recent years, there has been a renewed debate about how effective descriptions of the non-Fermi liquid state at a half-filled lowest Landau level ($\nu = 1/2$) of the two-dimensional electron gas might realize an emergent Landau level particle-hole (PH) symmetry \cite{PhysRevLett.50.1219, girvin1984}, found in electrical Hall transport \cite{Shahar1995, Wong1996, Pan2019} and numerical \cite{rezayi2000, Geraedtsetal2015} experiments.
The seminal theory of the half-filled Landau level of Halperin, Lee, and Read \cite{halperinleeread}, which has received substantial experimental support \cite{Willett97}, describes the $\nu=1/2$ state in terms of non-relativistic composite fermions in an effective magnetic field that vanishes at half-filling (see \cite{Jainbook, Fradkinbook} for pedagogical introductions).
However, the HLR theory appears to treat electrons and holes asymmetrically \cite{kivelson1997, BMF2015}.
For instance, it is naively unclear how composite fermions in zero effective magnetic field might produce the Hall effect $\sigma_{xy}^{\rm cf} = - {1 \over 4\pi}$ that PH symmetry requires \cite{kivelson1997}.
(We use the convention $k_B = c = \hbar = e = 1$.)

Two lines of thought point towards a possible resolution.
The first comes by way of an a priori different composite fermion theory, introduced by Son \cite{Son2015}.
In this Dirac composite fermion theory, the half-filled Landau level is described by a $(2+1)$-dimensional theory of quantum electrodynamics in which PH symmetry is a manifest invariance.
This theory is part of a larger web of $(2+1)$-dimensional quantum field theory dualities \cite{2018arXiv181005174S}. 
On the other hand, it has recently been shown that HLR mean-field theory {\it can} produce PH symmetric electrical response, if quenched disorder is properly included in the form of a precisely correlated random chemical potential and magnetic flux \cite{2017PhRvX...7c1029W, 2018arXiv180307767K, PhysRevB.98.115105}.
(Mean-field theory means that fluctuations of an emergent gauge field coupling to the composite fermion are ignored.)
Furthermore, both composite fermion theories yield identical predictions for a number of observables in mean-field theory \cite{Son2015, PhysRevLett.117.216403, 2017PhRvX...7c1029W, PhysRevB.95.235424, PhysRevB.99.205151}, e.g., thermopower at half-filling and magnetoroton spectra away from half-filling.
These results suggest that the HLR and Dirac composite fermion theories may belong to the same universality class.

To what extent do these results extend beyond the mean-field approximation? 
How do alternative experimental probes constrain the description of the $\nu=1/2$ state?
The aim of this paper is to address both of these questions within the Dirac composite fermion theory.
Prior work has identified observables that may possibly differ in the two composite fermion theories:
Son and Levin \cite{LevinSon2016} have derived a linear relation between the Hall conductivity and susceptibility that any PH symmetric theory must satisfy;
Wang and Senthil \cite{PhysRevB.94.245107} have determined how PH symmetry constrains the thermal Hall response of the HLR theory;
using the microscopic composite fermion wave function approach, Balram, Toke, and Jain \cite{BalramRifmmodeCsabaJain2015} found that Friedel oscillations in the pair-correlation function are symmetric about $\nu=1/2$.

\subsection{Weiss oscillations and the $\nu=1/2$ state}

Here, we study theoretically commensurability oscillations in the magnetoresistance near $\nu=1/2$, focusing on those oscillations that result from the presence of a periodic one-dimensional static potential \cite{Willett97}.
These commensurability oscillations are commonly known as Weiss oscillations \cite{Weissfirst, gerhardtsweissklitzing, winkler1989landau, Weiss1990}.
For a free two-dimensional Fermi gas, the locations of the Weiss oscillation minima, say, as a function of the transverse magnetic field $b$, satisfy
\begin{align}
\label{weissformulafreefermions}
\ell_{b}^2 = {d \over 2 k_F}\Big(p + \phi \Big),\quad p = 1, 2, 3, \ldots,
\end{align}
where $\ell_b = 1/\sqrt{|b|}$ is the magnetic length; $d$ is the period of the potential; $k_F$ is the Fermi wave vector; $\phi = +1/4$ for a periodic vector potential, while $\phi = -1/4$ for a periodic scalar potential \cite{peetersvasilopoulos1992scalar, zhanggerhardts}.
(Expressions for the oscillation minima when both potentials are present can be found in Refs.~\cite{peetersvasilopoulosmagnetic, gerhardts1996}.)

Early experiments \cite{Willett97} saw $p=1$ Weiss oscillation minima about $\nu=1/2$ due to an electrostatic {\it scalar} potential, upon identifying, in Eq.~\eqref{weissformulafreefermions}, $b = B - 4 \pi n_e$ with the effective magnetic field experienced by composite fermions ($B$ is the external magnetic field and $n_e$ is the electron density) and $k_F = \sqrt{4 \pi n_e}$ with the composite fermion Fermi wave vector, and choosing $\phi = + 1/4$.
These results, along with other commensurability oscillation experiments \cite{Willett97}, provided strong support for the general picture of the $\nu=1/2$ state suggested by the HLR theory.
In particular, the phenomenology near the $\nu=1/2$ state could be well described by an HLR mean-field theory in which composite fermions respond to an electronic scalar potential as a {\it vector} potential.

Recent improvements in sample quality and experimental design have allowed for an unprecedented refinement of these measurements.
Through a careful study of the oscillation minima corresponding to the first three harmonics ($p = 1, 2, 3$), Kamburov et al.~\cite{Kamburov2014} came to a remarkable conclusion that is in apparent disagreement with the above hypothesis (see \cite{shayeganreview2019} for a review of these and related experiments):
Weiss oscillation minima are well described by Eq.~\eqref{weissformulafreefermions} upon taking $k_F = \sqrt{4 \pi n_e}$ for $\nu < 1/2$, as before; but for $\nu > 1/2$, the inferred Fermi wave vector, $k_F = \sqrt{4 \pi ({B \over 2\pi} - n_e)}$, is determined by the density of holes.
In both cases, $\phi = +1/4$.
Might a theory of the $\nu=1/2$ state require two {\it different} composite fermion theories \cite{Kamburov2014, BMF2015}, a theory of composite electrons for $\nu < 1/2$ and a theory of composite holes for $\nu > 1/2$?  
If $k_F = \sqrt{4 \pi n_e}$ is instead taken for $1/2 < \nu < 1$, there is a roughly 2\% mismatch between the locations of the $p=1$ minimum obtained from Eq.~\eqref{weissformulafreefermions} and the nearest observed minimum; this discrepancy between theory and experiment decreases in magnitude as $p$ increases \cite{Kamburov2014}.
While the mismatch is small, it is systematic: it persists in a variety of different samples of varying mobilities and densities, as well as two-dimensional hole gases, which typically have larger effective masses (as well as near half-filling of other Landau levels \cite{shayeganreview2019}).
(This mismatch is the same magnitude as the difference between the electrical Hall conductivities produced by an HLR theory with $\sigma^{\rm cf}_{xy} = 0$ and an HLR theory with $\sigma^{\rm cf}_{xy} = -1/4\pi$, the composite fermion Hall conductivity required by PH symmetry; an equal value of the dissipative resistance \cite{Willett97} is assumed in both cases for this comparison. 
See Eq.~(48) of \cite{kivelson1997}.)

The hypothesis that composite fermions respond to an electric scalar potential as a purely magnetic one approximates HLR mean-field field theory. 
In fact, an electric scalar potential generates both a scalar and vector potential in the HLR theory.
(This observation by Wang et al.~\cite{2017PhRvX...7c1029W} is crucial for obtaining PH symmetric electrical Hall transport within HLR mean-field theory.)
However, the magnitude of the scalar potential is suppressed relative to the vector potential by a factor of $\ell_B/d \approx 1/50$ \cite{BMF2015}.
Cheung et al.~\cite{PhysRevB.95.235424} found that upon including the effects of the scalar potential in HLR mean-field theory, there is a slight correction to the expected locations of the oscillation minima {\it both} above and below $\nu=1/2$.
The nature of the corrections are such that HLR mean-field theories of composite electrons or composite holes that take either $k_F = \sqrt{4 \pi n_e}$ or $k_F = \sqrt{4 \pi ({B \over 2\pi} - n_e)}$ produce identical results.
In addition, the shifted oscillation minima are in agreement with the mean-field predictions of the Dirac composite fermion theory (at least within the regime of electronic parameters probed by experiment).
Unfortunately, the small disagreement between composite fermion mean-field theory and experiment persists, in this case for all values of $0 < \nu < 1$: for a given $p$, the observed oscillation minima are shifted inwards relative to the theoretical prediction by an amount that decreases as $\nu = 1/2$ is approached---see Fig.~\ref{weissoscillations}.

\subsection{Outline}

\begin{figure}
\center
\includegraphics[scale=0.47]{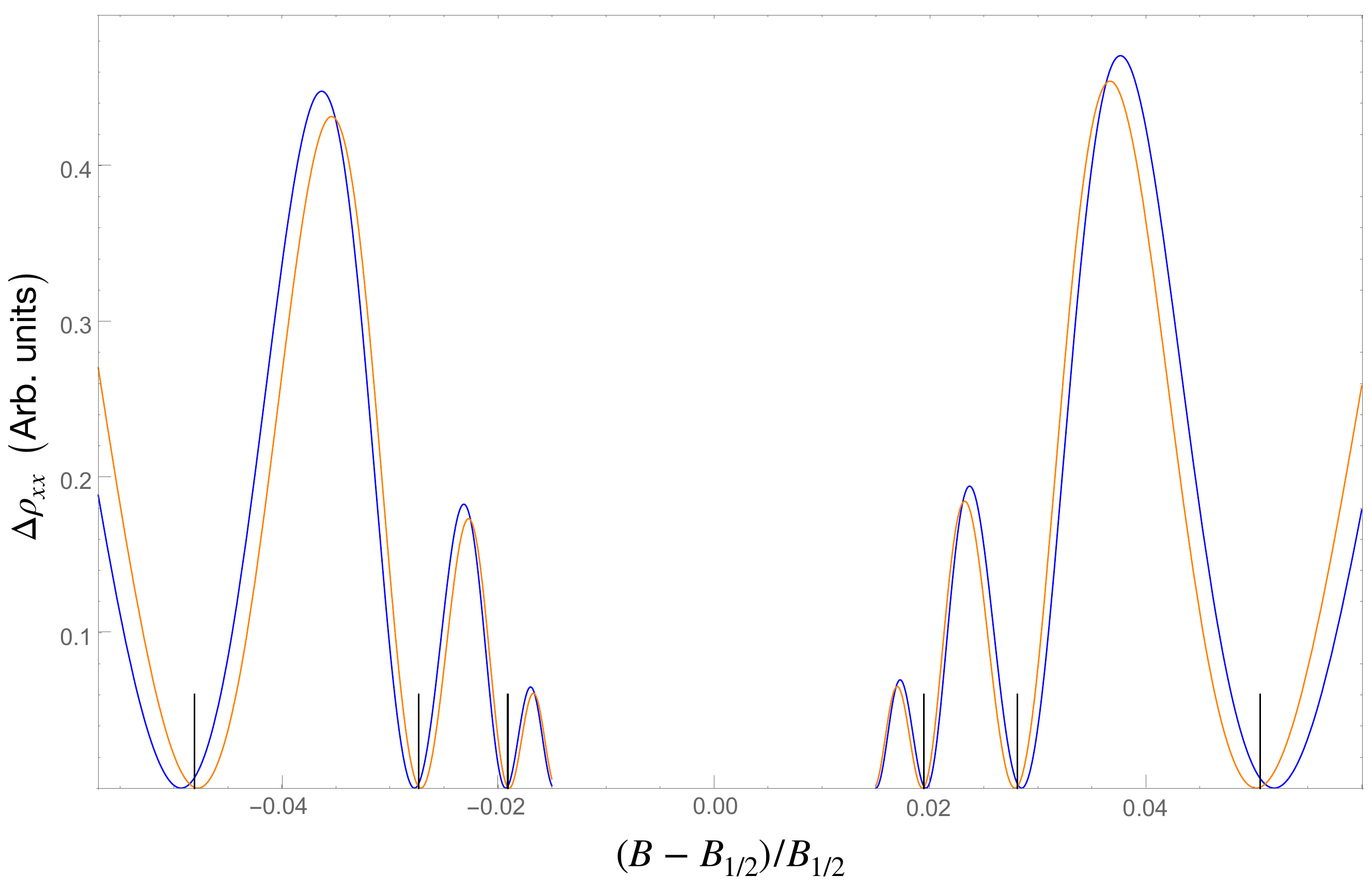}
\caption{Weiss oscillations of the Dirac composite fermion theory at fixed electron density $n_e$ and varying magnetic field $B$ about half-filling $B_{1/2}$ ($\ell_{B_{1/2}}/d = 0.03$ and $k_B T = 0.3\sqrt{2 B_{1/2}}$).
The blue curve corresponds to Dirac composite fermion mean-field theory \cite{PhysRevB.95.235424}. 
The orange curve includes the effects of a Dirac composite fermion mass $m \propto |B - 4\pi n_e|^{1/3} B^{1/6}$ induced by gauge fluctuations.
Vertical lines correspond to the observed oscillation minima \cite{Kamburov2014}.}
\label{weissoscillations}
 \end{figure}

In this paper, we consider the mismatch from the point of view of the Dirac composite fermion theory.
In perturbation theory about mean-field theory, we argue that the comparison with experiment can be improved if the effects of gauge field fluctuations are considered.
Our strategy is to include their effects by determining the fluctuation corrections to the mean-field Hamiltonian.
We obtain this corrected Hamiltonian through an approximate large $N$ flavor analysis of the Schwinger--Dyson equations \cite{Itzykson:1980rh} for the Dirac composite fermion theory.
The resulting Dirac composite fermion propagator specifies the input parameters, namely, the chemical potential and mass, of the corrected mean-field Hamiltonian.
We then follow the analysis by Cheung et al.~\cite{PhysRevB.95.235424} to determine the corrected Weiss oscillation curves.
Our results are summarized in Fig.~\ref{weissoscillations}.

To understand our results, it is helpful to reinterpret Eq.~\eqref{weissformulafreefermions} as a measure of a Dirac fermion density $n$ by replacing $k_F \mapsto \sqrt{4 \pi n}$ (we set the Fermi velocity to unity).
Any decrease in the density induces an inward shift of the Weiss oscillation minima determined by Eq.~\eqref{weissformulafreefermions} towards $b = 0$.
Dirac fermions of mass $m$, placed at chemical potential $\mu$ have a density $n = (\mu^2 - m^2)/4\pi$.
Our leading order analysis of the Schwinger--Dyson equations indicates that gauge fluctuations generate a mass $m$ away from $\nu=1/2$, while the chemical potential is unchanged.

Such dynamical mass generation in a non-zero magnetic field is known to occur in various (2+1)-dimensional theories of Dirac fermions (see \cite{Miransky:2015ava} for a review).
For example, in the theory of a free Dirac fermion at zero density, a uniform magnetic field sources a vacuum expectation value for the mass operator.
Short-ranged attractive interactions then induce a non-zero mass term in its effective Lagrangian \cite{Gusynin:1995nb}.
We show how a similar phenomenon occurs in the Dirac composite fermion theory.
This effect is also expected from the point of view of symmetry: PH symmetry forbids a Dirac composite fermion mass (see \S\ref{DiracCFreview}).
(Manifest PH symmetry is the essential advantage that the Dirac composite fermion theory confers to our analysis.)
Away from $\nu=1/2$, PH symmetry is broken and so all terms, consistent with the broken PH symmetry, are expected to be present in the effective Lagrangian. 
Note there is no symmetry preventing corrections to the Dirac composite fermion chemical potential; rather, it is found to be unaltered to leading order within our analysis.

We also comment upon the finite-temperature behavior of quantum oscillations near $\nu = 1/2$.
This behavior is interesting to consider because at finite temperatures, away from the long wavelength limit, differences in the HLR and Dirac composite fermion theories should appear.
We discuss how the temperature dependence of the Weiss oscillation amplitude might exhibit subtle differences between the two theories.

The remaining sections are organized as follows.
In \S\ref{DiracCFreview}, we review the Dirac composite fermion theory.
In \S\ref{SDsection}, we obtain an approximate solution to the Schwinger--Dyson equations.
In \S\ref{weisssection}, we use the chemical potential and mass of the resulting Dirac composite fermion propagator as input parameters for the ``fluctuation-improved" mean-field Hamiltonian and determine the resulting Weiss oscillations.
We discuss a few consequences of this analysis in \S\ref{discussion} and we conclude in \S\ref{conclusion}.
Appendix \ref{integralappendix} contains details of calculations summarized in the main text.

\section{Dirac composite fermions: review}
\label{DiracCFreview}

Electrons in the lowest Landau level near half-filling can be described by a Lagrangian of a 2-component Dirac electron $\Psi_e$ \cite{Son2015}:
\begin{align}
{\cal L}_e = \overline{\Psi}_e \gamma^\alpha (i \partial_\alpha + A_\alpha) \Psi_e - m_e \overline{\Psi}_e \Psi_e + {1 \over 8 \pi} \epsilon^{\alpha \beta \sigma} A_\alpha \partial_\beta A_\sigma + \ldots,
\end{align}
where $A_\alpha$ with $\alpha \in \{0, 1, 2 \}$ is the background electromagnetic gauge field, $\overline{\Psi}_e = \Psi_e^\dagger \gamma^0$, the $\gamma$ matrices $\gamma^0 = \sigma^3$, $\gamma^1 = i \sigma^1$, $\gamma^2 = i \sigma^2$ satisfy the Clifford algebra $\{\gamma^\alpha, \gamma^\beta\} = 2 \eta^{\alpha \beta}$ with $\eta^{\alpha \beta} = {\rm diag}(+1, -1, -1)$, and the anti-symmetric symbol $\epsilon^{0 1 2} = 1$. 
The benefit of the Dirac formulation is that the limit of infinite cyclotron energy $\omega_c = B/m_e$ can be smoothly achieved at fixed external magnetic field $B = \partial_1 A_2 - \partial_2 A_1 > 0$ by taking the electron mass $m_e \rightarrow 0$. 
The $\ldots$ include additional interactions, e.g., the Coulomb interaction and coupling to disorder.

The electron density,
\begin{align}
n_e = \Psi^\dagger_e \Psi_e + {B \over 4 \pi}.
\end{align}
Consequently, when $\nu \equiv {2\pi n_e/B} = 1/2$, the Dirac electrons half-fill the zeroth Landau level.
For $m_e = 0$ and $\nu = 1/2$, the Dirac Lagrangian is invariant under the anti-unitary ($i \mapsto - i$) PH transformation that takes $(t, x, y) \mapsto (-t, x, y)$,
\begin{align}
\label{electronPH}
\Psi_e & \mapsto - \gamma^0 \Psi_e^\ast, \cr
(A_0, A_1, A_2) & \mapsto (- A_0, A_1, A_2), 
\end{align}
and shifts the Lagrangian by a filled Landau level ${\cal L}_e \mapsto {\cal L}_e + {1 \over 4 \pi} \epsilon^{\alpha \beta \sigma} A_\alpha \partial_\beta A_\sigma$.

Son \cite{Son2015} conjectured that ${\cal L}_e$ is dual to the Dirac composite fermion Lagrangian,
\begin{align}
\label{CFlag}
{\cal L} = \overline{\psi} \gamma^\alpha (i \partial_\alpha + a_\alpha) \psi - m \overline{\psi} \psi - {1 \over 4 \pi} \epsilon^{\alpha \beta \sigma} a_\alpha \partial_\beta A_\sigma + {1 \over 8 \pi} \epsilon^{\alpha \beta \sigma} A_\alpha \partial_\beta A_\sigma - {1 \over 4 g^2} f_{\alpha \beta}^2 + \ldots,
\end{align}
where $\psi$ is the electrically-neutral Dirac composite fermion; $a_\alpha$ is a dynamical $U(1)$ gauge field with field strength $f_{\alpha \beta} = \partial_\alpha a_\beta - \partial_\beta a_\alpha$ and coupling $g$; and $m \propto m_e$ is the Dirac composite fermion mass.
$A_\alpha$ remains a non-dynamical gauge field, whose primary role in ${\cal L}$ is to determine how electromagnetism enters the Dirac composite fermion theory. 
As before, the $\ldots$ represent additional interactions, which can now involve the gauge field $a_\alpha$.
The duality between ${\cal L}_e$ and ${\cal L}$ obtains in the low-energy limit when $g \rightarrow \infty$.
See \cite{MetlitskiVishwanath2016, WangSenthilfirst2015, KMTW2015, Geraedtsetal2015, MrossAliceaMotrunichexplicitderivation2016, MurthyShankar2016halfull, Seiberg:2016gmd, PhysRevX.6.031043, 2018arXiv181111367S} for additional details about this duality and \cite{2018arXiv181005174S} for a recent review.

At weak coupling, the $a_0$ equation of motion implies the Dirac composite fermion density,
\begin{align}
\label{a0constraint}
\psi^\dagger \psi = {B \over 4 \pi}.
\end{align}
At strong coupling, the right-hand side of Eq.~\eqref{a0constraint} receives corrections from the $\ldots$ in ${\cal L}$ and should be replaced by $- {\delta {\cal L} \over \delta a_0} + \psi^\dagger \psi$.
In the Dirac composite fermion theory, the electron density,
\begin{align}
\label{edensityDiracCF}
n_e = {1 \over 4 \pi}(- b + B),
\end{align}
where the effective magnetic field $b = \partial_1 a_2 - \partial_2 a_1$.
In the Dirac composite fermion theory, the PH transformation takes $(t, x, y) \mapsto (-t, x, y)$,
\begin{align}
\label{DiracCFPH}
\psi & \mapsto \gamma^2 \psi, \cr
(a_0, a_1, a_2) & \mapsto (a_0, - a_1, - a_2), \cr
(A_0, A_1, A_2) & \mapsto (- A_0, A_1, A_2),
\end{align}  
and shifts the Lagrangian by a filled Landau level.
Intuitively, the PH transformation acts on the dynamical fields of ${\cal L}$ like a time-reversal transformation.
As such, PH symmetry requires $m=0$ and forbids a Chern-Simons term for $a_\alpha$.

Away from half-filling, PH symmetry is necessarily broken since Eq.~\eqref{edensityDiracCF} implies the effective magnetic field $b = B - 4 \pi n_e \neq 0$.
Consequently, we can no longer exclude any PH breaking term allowed by symmetry. 
In particular, we generally expect a Dirac mass to be induced by fluctuations.
Scaling implies the mass $m = \sqrt{B} f(\nu)$, where $f(\nu)$ is a scaling function of the filling fraction $\nu$. 
Unbroken PH symmetry at half-filling requires $f(\nu = 1/2) = 0$; 
away from $\nu=1/2$, it is possible that $m$ can have a non-trivial dependence on $B$ and $n_e$, as determined by $f(\nu)$.
In the next section, we study the Schwinger--Dyson equations to determine how fluctuations generate a mass $m$ away from $\nu = 1/2$ within an expansion where the number of Dirac composite fermion flavors $N \rightarrow \infty$.

\section{Dynamical mass generation in an effective magnetic field}
\label{SDsection}

Beginning with the works of Schwinger \cite{Schwinger:1951nm} and Ritus \cite{Ritus:1978cj}, there have been a number of studies on the effects of a background magnetic field on quantum electrodynamics in various dimensions.
In this paper, we rely most heavily on Refs.~\cite{Gorbar:2013upa, watson2014quark, 2015EPJC...75..167K}; see Ref.~\cite{Miransky:2015ava} for an excellent introduction to this formalism and for additional references.
We first summarize the relevant aspects of this formalism.
Then, we analyze the Schwinger--Dyson equations for the Dirac composite fermion theory away from half-filling when the fluctuations of the emergent gauge field $a_\alpha$ about a uniform $b \neq 0$ are considered.

\subsection{Dirac fermions in a magnetic field}
\label{Diracsinbfield}

At tree-level, i.e., in mean-field theory, the time-ordered real-space propagator $G_{0}(x,y)$ for a massive Dirac fermion in a uniform magnetic field  $(\overline{a}_0, \overline{a}_1, \overline{a}_2) = (0, 0, b x_1)$ can be written in the form,
\begin{align}
\label{treerealprop}
G_{0}(x,y) = e^{i \Phi(x,y)} \int {d^3 p \over (2\pi)^3} e^{i p_\alpha (x-y)^\alpha} G_{0}(p),
\end{align}
where the Schwinger phase,
\begin{align}
\Phi(x,y) = - {b \over 2} (x_2 - y_2) (x_1 + y_1).
\end{align}
The tree-level pseudo-momentum-space propagator,
\begin{align}
- i G_{0}(p) & = i \int_0^\infty d s e^{i s\Big((p_0 + \mu_0 + i \epsilon_{p_0})^2 - m_0^2 + i \delta - {p_1^2 + p_2^2 \over b s} \tan(b s)  \Big)} \cr
& \times \Big[(p_\alpha + \mu_0 \delta_{\alpha, 0})\gamma^\alpha - i b \Big((p_0 + \mu_0) \mathbb{I} + m_0 \gamma^0 \Big) \tan(b s) + p_i \gamma^i \tan^2(b s) \Big],
\end{align}
where the pseudo-momenta $p = (p_0, p_1, p_2)$ are analogous to the conserved momenta in a translationally-invariant system, $\mu_0$ is a chemical potential, $m_0$ is a mass, $\epsilon_{p_0} = {\rm sign}(p_0) \epsilon$ with the infinitesimal $\epsilon > 0$ ensures the Feynman pole prescription is satisfied, $\delta > 0$ is an infinitesimal included for convergence of the $s$ integral, and $\mathbb{I}$ is the $2 \times 2$ identity matrix.
Expanding in $b$:
\begin{align}
\label{treeprop}
- i G_{0}(p) & \equiv {(p_\alpha + \mu_0 \delta_{\alpha, 0}) \gamma^\alpha + m_0 \mathbb{I} \over (p_0 + \mu_0 + i \epsilon_{p_0})^2 - p_i^2 - m_0^2} + b {(p_0 + \mu_0) \mathbb{I} + m_0 \gamma^0 \over \Big((p_0 + \mu_0 + i \epsilon_{p_0})^2 - p_i^2 - m_0^2\Big)^2} + {\cal O}(b^2).
\end{align}
We imagine applying this formalism to the vicinity of $\nu=1/2$ when the effective magnetic field $b$ is small.
As such, we drop all ${\cal O}(b^2)$ and higher terms in the pseudo-momentum-space propagator.
For convenience, we use $G_{0}(p)$ to denote the linear expansion in Eq.~\eqref{treeprop} with higher order in $b$ terms excluded.
 
The tree-level inverse propagator $G^{-1}_{0}(x,y)$ satisfies 
\begin{align}
\int d^3 y\ G^{-1}_{0}(x,y) G_{0}(y, z) = \delta^{(3)}(x-z).
\end{align}
It takes a particularly simple form:
\begin{align}
\label{treepropinverse}
i G^{-1}_{0}(x,y) = e^{i \Phi(x,y)} \int {d^3 p \over (2\pi)^3} e^{i p_\alpha (x - y)^\alpha} \Big((p_\alpha + \mu_0 \delta_{\alpha, 0})\gamma^\alpha - m_0 \mathbb{I} \Big).
\end{align}
In contrast to $G_{0}(x,y)$, the magnetic field dependence is entirely parameterized by the Schwinger phase in $G^{-1}_{0}(x,y)$.

Both the propagator and its inverse are obtained after performing an infinite sum over all Landau levels.
Thus, $G_{0}(x,y)$ and $G_0^{-1}(x,y)$ in Eqs.~\eqref{treerealprop} and \eqref{treepropinverse} allow for a straightforward expansion about their translationally-invariant forms at $b=0$; see \cite{watson2014quark} for further discussion. 
In the Dirac composite fermion theory, $G_0^{-1}(x,y)$ defines the mean-field Lagrangian, from which the Hamiltonian readily follows; the Schwinger phase $\Phi(x,y)$ reminds us to include a non-zero magnetic field by the Peierls substitution.

We use the following ansatz for the exact real-space propagator:
\begin{align}
G(x,y) = e^{i \Phi(x,y)} \int {d^3 p \over (2 \pi)^3} e^{i p_\alpha (x-y)^\alpha} G(p).
\end{align}
For the exact pseudo-momentum propagator $G(p)$, we write
\begin{align}
\label{diracpropexpansion}
- i G(p) = - i G^{(0)}(p) - i G^{(1)}(p),
\end{align}
where
\begin{align}
\label{exactzero}
- i G^{(0)}(p) & = {\Big(p_\alpha + \mu_0 \delta_{\alpha, 0} - \Sigma_\alpha(p)\Big) \gamma^\alpha + \Sigma_m(p) \mathbb{I} \over (p_0 + \mu_0 - \Sigma_0(p) + i \epsilon_{p_0})^2 - (p_i - \Sigma_i(p))^2 - \Sigma_m^2(p)}, \\
\label{exactone}
- i G^{(1)}(p) & = b {\Big(p_0 + \mu_0 - \Sigma_0(p)\Big) \mathbb{I} + \Sigma_m(p) \gamma^0 \over \Big((p_0 + \mu_0 - \Sigma_0(p) + i \epsilon_{p_0})^2 - (p_i - \Sigma_i(p))^2 - \Sigma_m^2(p)\Big)^2}.
\end{align}
In contrast to the tree-level pseudo-momentum propagator, $G_0(p)$, both $G^{(0)}(p)$ and $G^{(1)}(p)$ are expected to depend on $b$ through the self-energies $\Sigma_m(p)$ and $\Sigma_\alpha(p)$, in addition to the explicit linear dependence that appears in $G^{(1)}(p)$. 
We write the exact inverse propagator as
\begin{align}
\label{exactinversedirac}
i G^{-1}(x,y) = e^{i \Phi(x,y)} \int {d^3 p \over (2 \pi)^3} e^{i p_\alpha (x - y)^\alpha} \left( \left( p_\alpha + \mu_0 \delta_{\alpha, 0} - \Sigma_\alpha \left( p \right) \right) \gamma^\alpha - \Sigma_m \left( p \right) \mathbb{I} \right).
\end{align}
In $G(p)$ and $G^{-1}(p)$, we set the tree-level mass $m_0 = 0$; this is consistent with the assumption of unbroken PH symmetry at $\nu = 1/2$.
The ansatze for the exact propagator and its inverse are simplifications of that which symmetry allows for a Dirac fermion in a magnetic field \cite{watson2014quark}.
Nevertheless, our ansatze are consistent to leading order in a $1/N$ analysis of the Schwinger--Dyson equations described in the next section.

In general, the self-energies $\Sigma_m(p)$ and $\Sigma_\alpha(p)$ are non-trivial functions of the pseudo-momenta $p$. 
We expect the low-energy dynamics of the fermions to be dominated by fluctuations about the Fermi surface.
Thus, we replace the self-energies as follows:
\begin{align}
\label{massselfenergy}
\Sigma_m(p_{\rm FS} + \delta p) & \mapsto \Sigma_m(p_{\rm FS}), \\
\label{momentaselfenergy}
\Sigma_\alpha(p_{\rm FS} + \delta p) & \mapsto \delta_{0 \alpha} \Sigma_0(p_{\rm FS}) + \delta p_\alpha \Sigma'_\alpha(p_{\rm FS}),
\end{align}
where $p_{\rm FS} = (0, p_i)$ lies on the Fermi surface (in mean-field theory, this is defined by $p_i^2 = \mu_0^2$ and $p_0 = 0$), $|\delta p_\alpha| \ll \mu_0$, $\Sigma'_{\alpha}(p_{\rm FS}) = \partial_{p_\alpha} \Sigma_\alpha(p = p_{\rm FS})$, and there is no sum over $\alpha$ in Eq.~\eqref{momentaselfenergy}.

$G^{-1}(x,y)$ determines the ``fluctuation-corrected" Dirac composite fermion mean-field Hamiltonian. 
The tree-level chemical potential and mass are corrected by the fermion self-energies $\Sigma_\alpha$ and $\Sigma_m$.
We define the physical mass,
\begin{align}
\label{dynamicalmass}
m = {\Sigma_m(p_{\rm FS}) \over 1 - \Sigma'_0(p_{\rm FS})} \equiv {\Sigma_m \over 1 - \Sigma'_0},
\end{align}
and chemical potential,
\begin{align}
\label{correctedchemicalpotential}
\mu = {\mu_0 - \Sigma_0 \over 1 - \Sigma'_0}.
\end{align}
The Schwinger phase $\Phi(x,y)$ in $G^{-1}(x,y)$ reminds us to to include the effective magnetic field $b$ via the Peierls substitution.

\subsection{Schwinger--Dyson equations: setup}

The Schwinger--Dyson equations \cite{Itzykson:1980rh} are a set of coupled integral equations that relate the exact fermion and gauge field propagators to one another by way of the exact cubic interaction vertex $\Gamma^\alpha$ coupling the Dirac composite fermion current to $a_\alpha$.
We will not solve the equations exactly; rather, we seek an approximate solution that one obtains within a large flavor generalization of the Dirac composite fermion theory.
We hope this approximate solution reflects a qualitative behavior of the Dirac composite fermion theory.

Specifically, we consider the Lagrangian,
\begin{align}
\label{largeNgeneneral}
{\cal L}_N = \overline{\psi}_n \gamma^\alpha (i \partial_\alpha + a_\alpha) \psi_n - {N \over 4 \pi} \epsilon^{\alpha \beta \sigma} a_\alpha \partial_\beta A_\sigma + {N \over 8 \pi} \epsilon^{\alpha \beta \sigma} A_\alpha \partial_\beta A_\sigma - {1 \over 4 g^2} f_{\alpha \beta}^2, 
\end{align}
where the different fermion flavors are labeled by $n = 1, \ldots, N$.
When $N = 1$, we recover the Dirac composite fermion theory.
In ${\cal L}_N$,  $n_e = {\delta {\cal L}_N/\delta A_0} = {N \over 4 \pi} (B - b)$; thus, in our large $N$ theory, half-filling means $\nu=N/2$.
To make contact with the formalism of \S\ref{Diracsinbfield}, we introduce a $SU(N)$-invariant chemical potential $\mu_0 = \sqrt{B}$ and we factor out the uniform effective magnetic field $(\overline{a}_0, \overline{a}_1, \overline{a}_2) = (0, 0, b x_1)$ that is generated away from half-filling from the dynamical fluctuations of the emergent gauge field $a_\alpha$.
Setting $A_\alpha = 0$, Eq.~\eqref{largeNgeneneral} becomes
\begin{align}
\label{largeNgeneneralconstrained}
{\cal L}_N = \overline{\psi}_n \gamma^\alpha (i \partial_\alpha + \overline{a}_\alpha + a_\alpha) \psi_n + \mu_0 \psi^\dagger_n \psi_n - {1 \over 4 g^2} f_{\alpha \beta}^2.
\end{align}
This is the large $N$ theory that we analyze.

To leading order in $N$, the Ward identity implies that there are no corrections to the cubic interaction vertex at $\nu = 1/2$ \cite{2018arXiv180802140R}.\footnote{Furthermore, there are no corrections to this vertex if the Dirac composite fermion is given a non-zero bare mass $m_0^2 \ll \mu_0^2$ at $b=0$. We thank N. Rombes and S. Chakravarty for correspondence on this point.} 
Taking $\Gamma^\alpha = \gamma^\alpha$, the Schwinger--Dyson equations for ${\cal L}_N$ become:
\begin{align}
\label{realspaceSDfermion}
i G^{-1}(x,y) - i G^{-1}_{0}(x,y) & = \gamma^\alpha G(x,y) \gamma^\beta \Pi^{-1}_{\alpha \beta}(x - y), \\
\label{realspaceSDgauge}
i \Pi^{\alpha \beta}(x-y) - i \Pi^{\alpha \beta}_0(x-y) & = N {\rm tr}\Big[ \gamma^\alpha G(x,y) \gamma^\beta G(y,x) \Big],
\end{align}
where $\Pi^{\alpha \beta}(x-y)$ is the gauge field self-energy, $\Pi_0^{\alpha \beta}(x-y)$ is the kinetic term for $a_\alpha$ contributed by its Maxwell term, and we have taken the fermion propagator $G_{n, n'}(x,y) = G(x,y) \delta_{n, n'}$ to be diagonal in flavor space.
$G(x,y)$ and $G_0(x,y)$ are defined in Eqs.~\eqref{diracpropexpansion} and \eqref{treeprop}.
The factor of $N$ in Eq.~\eqref{realspaceSDgauge} arises from the $N$ flavors in the fermion loop.

Upon substituting the Fourier transform $\Pi^{\alpha \beta}(p)$, defined by
\begin{align}
\Pi^{\alpha \beta}(x-y) = \int {d^3 p \over (2 \pi)^3} e^{i p_\sigma (x-y)^\sigma} \Pi^{\alpha \beta}(p),
\end{align}
and Eqs~\eqref{treepropinverse}, \eqref{diracpropexpansion}, and \eqref{exactinversedirac} into the Schwinger--Dyson equations, \eqref{realspaceSDfermion} and \eqref{realspaceSDgauge} become \cite{watson2014quark}
\begin{align}
\label{SDfermion}
i \Sigma_\alpha(q) \gamma^\alpha + i \Sigma_m(q) \mathbb{I} = \int {d^3 p \over (2 \pi)^3} \gamma^{\alpha} G(p+q) \gamma^\beta \Pi^{-1}_{\alpha \beta}(p), \\
\label{SDgauge}
i \Pi^{\alpha \beta}(\delta q) = N \int {d^3 p \over (2 \pi)^3} {\rm tr}\Big[\gamma^\alpha G(p) \gamma^\beta G(p + \delta q) \Big],
\end{align}
where $q = q_{\rm FS} + \delta q$.
We aim to solve these equations.

Our ansatz for the fermion self-energies is motivated by similar studies of $(2+1)$-dimensional quantum electrodynamics at zero density \cite{PhysRevD.29.2423, AppelquistNashWijewardhanaQED3, PhysRevLett.62.3024}.
We consider the $1/N$ expansion for the fermion self-energies,
\begin{align}
\label{explicitexpansion}
\Sigma_\alpha & = \Sigma_\alpha^{(1)} + \Sigma_\alpha^{(2)} + \ldots, \cr
\Sigma_m & = \Sigma_m^{(1)} + \Sigma_m^{(2)} + \ldots.
\end{align}
All terms and all ratios of successive terms in Eq.~\eqref{explicitexpansion} vanish as $N\rightarrow \infty$.
Ignoring terms with $i \geq 2$, we set $\Sigma_\alpha = \Sigma_\alpha^{(1)} = 0$ and $\Sigma_m = \Sigma_m^{(1)}$, and find a self-consistent solution to the Schwinger--Dyson equation in terms of $\Sigma_m^{(1)}$ and $\Pi^{\alpha \beta}$. 
This choice is consistent with the Ward identity, to leading order in $1/N$.
From Eqs.~\eqref{dynamicalmass} and \eqref{correctedchemicalpotential}, the resulting solution implies $m = \Sigma_m^{(1)}$ and $\mu = \mu_0$ to leading order in $1/N$.
We then calculate the leading perturbative correction $\Sigma^{(2)}_\alpha$ to $\Sigma_\alpha$ and verify that $\Sigma^{(2)}_\alpha/\Sigma^{(1)}_m \rightarrow 0$ as $N \rightarrow \infty$.

\subsection{Gauge field self-energy}
\label{gaugeselfenergy}

The gauge field self-energy factorizes into PH symmetry even and odd parts:
\begin{align}
\label{gaugeinvariance}
\Pi^{\alpha \beta}(q) = \Pi^{\alpha \beta}_{\rm even}(q) + \Pi^{\alpha \beta}_{\rm odd}(q).
\end{align}
As the PH transformation acts like time-reversal, $\Pi^{\alpha \beta}_{\rm even}(q)$ contains the Maxwell term for $a_\alpha$, while $\Pi^{\alpha \beta}_{\rm odd}(q)$---which can only be non-zero when PH symmetry is broken---can contain a Chern-Simons term for $a_\alpha$.

To leading order in $b$, we substitute $G(p) = G^{(0)}(p)$ into Eq.~\eqref{SDgauge} and first compute
\begin{align}
\Pi^{\alpha \beta}_{\rm odd}(\delta q) = i \epsilon^{\alpha \beta \sigma} \delta q_\sigma \Pi_{\rm odd}(\delta q) = - i N \Big\{ \int {d^3 p \over (2\pi)^3} {\rm tr}\Big[ \gamma^\alpha G^{(0)}(p) \gamma^\beta G^{(0)}(p + \delta q)\Big] \Big\}_{\rm odd},
\end{align}
where $\{ \cdot \}_{\rm odd}$ indicates the PH odd term is isolated.
We find
\begin{align}
\label{inducedCS}
\Pi_{\rm odd}(0) 
& = {N \over 4 \pi} \Big(\Theta(|\Sigma_m| - \mu_0) {\Sigma_m \over |\Sigma_m|} + \Theta(\mu_0 - |\Sigma_m|) {\Sigma_m \over \mu_0} \Big),
\end{align}
where $\Theta(x)$ is the step function.
See Appendix \ref{gaugefieldselfenergyappendix} for details.
Additional momentum dependence in $\Pi_{\rm odd}(q)$ is subdominant at low energies.
For $\mu_0 > |\Sigma_m|$, Eq.~\eqref{inducedCS} implies an effective Chern-Simons term for $\alpha_\alpha$ with level,
\begin{align}
\label{cslevel}
k = {N \over 2} {\Sigma_m \over \mu_0},
\end{align}
is generated if $\Sigma_m \neq 0$.
(This non-quantized Chern-Simons level is reminiscent of the anomalous Hall effect \cite{PhysRevLett.93.206602}.)

Next, consider 
\begin{align}
\Pi^{\alpha \beta}_{\rm even}(\delta q) - \Pi^{\alpha \beta}_{0}(\delta q) = - i N \Big\{ \int {d^3 p \over (2\pi)^3} {\rm tr}\Big[ \gamma^\alpha G^{(0)}(p) \gamma^\beta G^{(0)}(p + \delta q)\Big] \Big\}_{\rm even},
\end{align}
where $\{ \cdot \}_{\rm even}$ indicates the PH even term is isolated and we have again substituted $G(p) = G^{(0)}(p)$. 
The Maxwell kinetic term is
\begin{align}
\Pi^{\alpha \beta}_{0}(q) = q^2 \eta^{\alpha \beta} - q^\alpha q^\beta.
\end{align}
Ref.~\cite{Miransky:2001qs} finds:
\begin{align}
\Pi^{00}_{\rm even}(q_0, q_i) - \Pi^{0 0}_{0}(q) & = \Pi_l(q_0, q_i), \cr
\Pi^{0 i}_{\rm even}(q_0, q_i) - \Pi^{0 i}_{0}(q) & = q_0 {q^i \over q_i^2} \Pi_l(q_0, q_i), \cr
\Pi^{i j}_{\rm even}(q_0, q_i) - \Pi^{0 i}_{0}(q) & = (\delta^{i j} - {q^i q^j \over q_k^2}) \Pi_t(q_0, q_i) + {q_0^2 q^i q^j \over (q_k^2)^2} \Pi_l(q_0, q_i),
\end{align}
where
\begin{align}
\Pi_l(q_0, q_i) & = \mu_0 N \Big(\sqrt{{q_0^2 \over q_0^2 - q_i^2}} - 1 \Big), \cr
\Pi_t(q_0, q_i) & = \mu_0 N - {q_0^2 - q_i^2 \over q_k^2} \Pi_l(q_0, q_i).
\end{align}
We have simplified the expressions for $\Pi_l$ and $\Pi_t$ by taking $q_0^2 - q_i^2 > 0$ and by setting the common proportionality constant to unity.
The precise behaviors of $\Pi_l$ and $\Pi_t$ and their effects on $a_\alpha$ depend upon whether $|q_0| < |q_i|$ or $|q_i| < |q_0|$.
For instance, when $|q_0| < |q_i|$ (small frequency transfers, but potentially large $\sim 2k_F$ momenta transfers) and in the absence of $\Pi_{\rm odd}^{\alpha \beta}$, $\Pi_l$ gives rise to the usual Debye screening of the ``electric" component of $a_\alpha$ and $\Pi_t$ results in the Landau damping of the ``magnetic" component of $a_\alpha$ \cite{Miransky:2001qs}, familiar from Fermi liquid theory \cite{PhysRevB.8.2649}.
These corrections dominate the tree-level Maxwell term for $a_\alpha$ at low energies.

In our analysis of the fermion self-energy in the next section, we focus on the regime $|q_i| \leq |q_0|$.
In this case, $\Pi_l$ and $\Pi_t$ provide non-singular corrections to the Maxwell term for $a_\alpha$ and will be ignored.
At low energies, $g \rightarrow \infty$, the effects of the Maxwell term are suppressed compared with the Chern-Simons term \cite{1999tald.conf..177D}.
Thus, to find the effective gauge field propagator $\Pi^{-1}_{\alpha \beta}(q)$ for use in Eq.~\eqref{SDfermion}, we drop $\Pi_{\rm even}^{\alpha \beta}(q)$, add the covariant gauge fixing term  $- {1 \over 2 \xi} q^\alpha q^\beta$ to $\Pi^{\alpha \beta}_{\rm odd}(q)$, and invert. 
Choosing Feynman gauge $\xi = 0$, we obtain:
\begin{align}
\label{gaugepropagator}
\Pi_{\alpha \beta}^{-1}(q) = {2 \pi \over k} {\epsilon_{\alpha \beta \sigma} q^\sigma \over q^2},
\end{align}
where $k$ is given in Eq.~\eqref{cslevel}.
It is with this gauge field propagator that we find a self-consistent solution to the Schwinger--Dyson equation for the fermion self-energy $\Sigma_m$ in \S\ref{fermionselfenergy}.

Instantaneous density-density interactions between electrons give rise to additional gauge field kinetic terms in ${\cal L}$.
Such terms, which should therefore be included in the tree-level Lagrangian ${\cal L}_N$, generally contribute to $\Pi_0^{\alpha \beta} \subset \Pi_{\rm even}^{\alpha \beta}$.
To understand their possible effects in the kinematic regime $|q_i| \leq |q_0|$, we set $a_0 = 0$ and decompose the spatial components of the gauge field in terms of its longitudinal and transverse modes:
\begin{align} 
a_i(q) = - i \hat{q}_i a_L(q) - i \epsilon_{j i} \hat{q}_j a_T(q),
\end{align}
where the normalized spatial momenta $\hat{q}_i = q_i/|\vec{q}\,|$.
An un-screened Coulomb interaction dualizes to a term in ${\cal L}$ proportional to $|\vec{q}\,|^{z-1} a_T(-q) a_T(q)$ with $z=2$; a short-ranged interaction give $z=3$ (see Sec.~3.4 of \cite{KMTW2015}).
(We are working in momentum space for this analysis.)
On the other hand, the effective Chern-Simons term is proportional to $i q_0 a_L(-q) a_T(q)$; there is no $a_L - a_L$ or $a_T - a_T$ Chern-Simons coupling.
We consider $z > 2$ in our analysis below. 
In this regime, the effects of any such screened interaction are expected to be subdominant compared with those of the Chern-Simons term, as such interactions correspond to higher-order terms in the derivative expansion.

\subsection{Fermion self-energy}
\label{fermionselfenergy}

We now study Eq.~\eqref{SDfermion} for the $\Sigma_m$ and $\Sigma_0$ components of the Dirac composite fermion self-energy using the effective gauge field propagator in Eq.~\eqref{gaugepropagator}.

\subsubsection{$\Sigma_m$}

Taking the trace of both sides of Eq.~\eqref{SDfermion} and setting $\delta q_\alpha = 0$, we find:
\begin{align}
\label{massselfenergystart}
i \Sigma_m(q_{\rm FS}) & = i {\cal M}^{(0)}(q_{\rm FS}) + i {\cal M}^{(1)}(q_{\rm FS}),
\end{align}
where 
\begin{align}
i {\cal M}^{(0)}(q_{\rm FS}) & = {1 \over 2} \int {d^3 p \over (2\pi)^3} {\rm tr} \Big[ \gamma^\alpha G^{(0)}(p + q_{\rm FS}) \gamma^\beta \Big({2 \pi \over k} {\epsilon_{\alpha \beta \sigma} p^\sigma \over p^2} \Big) \Big], \\
i {\cal M}^{(1)}(q_{\rm FS}) & = {1 \over 2} \int {d^3 p \over (2\pi)^3} {\rm tr} \Big[ \gamma^\alpha G^{(1)}(p + q_{\rm FS}) \gamma^\beta \Big({2 \pi \over k} {\epsilon_{\alpha \beta \sigma} p^\sigma \over p^2} \Big) \Big],
\end{align}
and $G^{(0}(p)$ and $G^{(1)}(p)$ are given in Eqs.~\eqref{exactzero} and \eqref{exactone}.
Recall that we set $\Sigma_\alpha = 0$ and only retain $\Sigma_m$ when using $G^{(0)}(p)$ and $G^{(1)}(p)$ to evaluate ${\cal M}^{(0)}$ and ${\cal M}^{(1)}$.
The details of our evaluation of ${\cal M}^{(0)}$ and ${\cal M}^{(1)}$ are given in Appendix \ref{fermionselfenergyappendix}.
Here, we quote the results:
\begin{align}
{\cal M}^{(0)} & = - {2 \mu_0 {\rm sign}(\Sigma_m) \over N}, \\
{\cal M}^{(1)} & = {2 \over 3} {b \mu_0^2 \over N |\Sigma_m|^3}.
\end{align}

Thus, $\Sigma_m$ solves:
\begin{align}
\label{massselfequation}
\Sigma_m = - {2 \mu_0 {\rm sign}(\Sigma_m) \over N} +  {2 \over 3} {b \mu_0^2 \over N |\Sigma_m|^3}.
\end{align}
When $b=0$, the only solution is $\Sigma_m = 0$, consistent with our expectation that PH symmetry is unbroken at $\nu=1/2$.
Dimensional analysis and $1/N$ scaling implies
\begin{align}
\Sigma_m = {\mu_0 \over N} f\Big({b N^3 \over \mu_0^2}\Big).
\end{align}
We find that $\Sigma_m$ has the following asymptotics: for fixed $|b|/\mu_0^2 \approx 10^{-1}$,
\begin{align}
\label{sigmalargeN}
\Sigma_m & = \mu_0 {\rm sign}(b) \Big({|b| \over \mu_0^2 N} \Big)^{1/4} \Big[c_1 + c_2 \Big({\mu_0^2\over |b| N^3}\Big)^{1/4}  + \ldots \Big],
\end{align}
where $c_1 \approx 0.9$, $c_2 \approx -0.5$, and the $\ldots$ are suppressed as $N \rightarrow \infty$;
while for fixed $N$,
\begin{align}
\label{sigmasmallb}
\Sigma_m = \mu_0 {\rm sign}(b) \Big({|b| \over \mu_0^2}\Big)^{1/3}\Big[c_3 + c_4 \Big({|b| N^3 \over \mu_0^2}\Big)^{1/3} + \ldots \Big],
\end{align}
where $c_3 \approx 0.69$, $c_4 \approx -0.08$, and the $\ldots$ vanish as $|b|/\mu_0^2 \rightarrow 0$.

\subsubsection{$\Sigma_0$}

We now consider the leading perturbative correction to $\Sigma_0$.
This allows us to calculate the corrections to $\Sigma'_0$ and the chemical potential $\mu_0$.

To evaluate the leading correction to $\Sigma_0$ that one obtains when $G(p) = G^{(0)}(p)$, we multiply both sides of Eq.~\eqref{SDfermion} by $\gamma^0$ on the left and take the trace to find:
\begin{align}
i \Sigma_0(q)  = {1 \over 2} \int {d^3 p \over (2\pi)^3} {\rm tr} \Big[\gamma^0 \gamma^\alpha G^{(0)}(p + q) \gamma^\beta \Big({2 \pi \over k} {\epsilon_{\alpha \beta \sigma} p^\sigma \over p^2} \Big) \Big],
\end{align}
where $q^\alpha = q^\alpha_{\rm FS} + q_0 \delta^{\alpha 0}$.
As detailed in Appendix \ref{fermionselfenergyappendix}, we find the leading correction $\Sigma_0^{(2)}$ to $\Sigma_0$ (see Eq.~\eqref{explicitexpansion}) for $|q_0|/\mu_0 \ll \Sigma^2_m/\mu_0^2$,
\begin{align}
\label{wfrenorm}
i \Sigma_0^{(2)}(q_{\rm FS}) = - i {2 \mu_0 \over 3 N |\Sigma_m|} (q_0 + \mu_0).
\end{align}
At large $N$, we use Eq.~\eqref{sigmalargeN} for $\Sigma_m$ to find $\Sigma_0 \propto \Sigma'_0 \propto N^{-3/4}$.
This vanishes by a factor of $N^{-1/2}$ {\it faster} than $\Sigma_m$ and so it is relatively suppressed as $N \rightarrow \infty$.
Next-order terms in $\Sigma_\alpha$ and $\Sigma_m$ are obtained by self-consistently solving the Schwinger--Dyson equations with propagators corrected by the leading self-energy corrections.
We have checked that the other components of $\Sigma_\alpha$ are likewise suppressed at large $N$; as such and because they do not enter our subsequent calculations, we will not discuss them further.
Because $\Sigma_m$ vanishes at half-filling, we may only ignore $\Sigma'_0$ for sufficiently large $|b|/\mu_0^2$ at large $N$.

\subsubsection{Dynamically-generated mass and corrected chemical potential}

We are now ready to evaluate Eq.~\eqref{dynamicalmass} for the dynamically-generated mass.
We extrapolate our large $N$ solution for $\Sigma_m$ to $N=1$ using Eq.~\eqref{sigmasmallb}:
\begin{align}
\label{mass}
m = {\Sigma^{(1)}_m \over 1 - \Sigma'^{(1)}_0} \approx .69 {\rm sign}(b) |b|^{1/3} B^{1/6},
\end{align}
where we set $\mu_0 = \sqrt{B}$.
The specific behavior of the mass $m$, away from $\nu=1/2$, depends on whether the electron density $n_e$ or external magnetic field $B$ is fixed.
At fixed $B$, the magnitude of $m$ is symmetric as function of $n_e$ about half-filling; on the other hand, $|m|$ is asymmetric for fixed $n_e$ and varying $B$.
Using Eqs.~\eqref{correctedchemicalpotential} and \eqref{explicitexpansion}, the chemical potential,
\begin{align}
\mu = {\mu_0 - \Sigma^{(1)}_0 \over 1 - \Sigma'^{(1)}_0} = \sqrt{B}.
\end{align}
These results imply that the Dirac composite fermion density and mass are corrected in such a way that the chemical potential is unaffected.

In our analysis of the Weiss oscillations in the next section, we ignore all higher-order in $1/N$ corrections and assume that a mass term is the dominant correction to the Dirac composite fermion mean-field Hamiltonian away from $\nu=1/2$.
The chemical potential for this fluctuation-improved mean-field Hamiltonian will be taken to be $\mu = \sqrt{B}$.

\section{Weiss oscillations of massive Dirac composite fermions}
\label{weisssection}

Following earlier work \cite{matulispeetersweiss2007, tahirsabeehmagneticweiss2008, 2016arXiv161004068B, PhysRevB.95.235424}, we now study the effect of the field-dependent mass of Eq.~\eqref{mass} on the Weiss oscillations near $\nu=1/2$ using the fluctuation-improved Dirac composite fermion mean-field theory.
We find that a non-zero mass results in an inward shift of the locations of the oscillation minima toward half-filling.

\subsection{Setup}

We are interested in determining the quantum oscillations in the electrical resistivity near $\nu=1/2$ that result from a one-dimensional periodic scalar potential.
In the Dirac composite fermion theory, the dc electrical conductivity,
\begin{align}
\label{conductivitydictionary}
\sigma_{ij} = {1 \over 4\pi} \Big(\epsilon_{ij} - {1 \over 2} \epsilon_{i k} (\sigma^\psi)_{kl}^{-1} \epsilon_{l j} \Big),
\end{align}
where the (dimensionless) dc Dirac composite fermion conductivity.
This equality is true at weak coupling; at strong coupling, $\langle \overline{\psi}\gamma_i \psi(- q_0) \overline{\psi}\gamma_j \psi(q_0)  \rangle$ should be replaced by the exact gauge field $a_\alpha$ self-energy, evaluated at $q_1 = q_2 = 0$.
\begin{align}
\sigma^\psi_{ij} = \lim_{q_0 \rightarrow 0} {\langle \overline{\psi}\gamma_i \psi(- q_0) \overline{\psi}\gamma_j \psi(q_0)  \rangle \over i q_0}.
\end{align}
Thus, the longitudinal electrical resistivity,
\begin{align}
\label{resistivity}
\rho_{i i} \propto |\epsilon_{i j}| \sigma_{j j}^\psi,
\end{align}
where there is no sum over repeated indices.
When a one-dimensional periodic scalar potential, $A_0 = V \cos(K x_1)$ with $K = 2\pi/d$, is applied to the electronic system, the $a_2$ equation of motion following from the Dirac composite fermion Lagrangian \eqref{CFlag} implies
\begin{align}
\overline{\psi} \gamma^2 \psi = - {K V \over 4\pi} \sin(K x).
\end{align}
We accommodate this current modulation within Dirac composite fermion mean-field theory by turning on a modulated perturbation to the emergent vector potential,
\begin{align}
\label{backgroundvector}
\delta \vec{a} = \Big(0, W \sin(K x_1)\Big),
\end{align}
where $W = W(V)$ vanishes when $V = 0$.
(Fluctuations will also generate a modulation in the Dirac composite fermion chemical potential; we ignore such effects here.)
Putting together Eqs.~\eqref{resistivity} and \eqref{backgroundvector}, our goal in this section is to determine the correction to $\sigma_{jj}^\psi$ due to $\delta \vec{a}$,
\begin{align}
\label{dictionaryoscillation}
\Delta \rho_{ii} \propto |\epsilon_{ij}| \Delta \sigma^\psi_{jj}.
\end{align}

In Dirac composite fermion mean-field theory, corrected by Eq.~\eqref{mass}, the calculation of $\Delta \sigma_{ij}^\psi$ simplifies to the determination of the conductivity of a free {\it massive} Dirac fermion.
We use the Kubo formula \cite{charbonneauvVV1982} to find the conductivity correction:
\begin{align}
\label{kubo}
\Delta \sigma_{ij}^\psi = {1 \over L_1 L_2} \Sigma_M \Big( \partial_{E_M} f_D(E_M) \Big) \tau(E_M) v_i^M v_j^M,
\end{align}
where $L_1$ ($L_2$) is the length of the system in the $x_1$-direction ($x_2$-direction), $\beta^{-1} = T$ is the temperature, $M$ denotes the quantum numbers of the single-particle states, $f^{-1}_D(E) = 1 + \exp(\beta(E - \mu))$ is the Fermi-Dirac distribution function with chemical potential $\mu = \sqrt{B}$, $\tau(E_M)$ is the scattering time for states at energy $E_M$, and $v_i^M = \partial_{p_i} E_M$ is the velocity correction in the $x_i$-direction of the state $M$ due to the periodic vector potential. 
As before, the Fermi velocity is set to unity.
Assuming constant $\tau(E) = \tau \neq 0$, we only need to calculate how the energies $E_M$ are affected by $\delta \vec{a}$, which in turn will determine the velocities $v_i^M$.
We will show that the leading correction in $W$ to $E_M$ only contributes to $v_2^M$.
Calling $x_1 = x$ and $x_2 = y$, this implies the dominant correction is to $\Delta \rho_{xx} \propto \Delta \sigma_{yy}^\psi$.
There are generally oscillatory corrections to $\rho_{yy}$ and $\rho_{xy}$, however, their amplitudes are typically less prominent and so we concentrate on $\Delta \rho_{xx}$ here.

\subsection{Dirac composite fermion Weiss oscillations}

The Dirac composite fermion mean-field Hamiltonian, corrected by Eq.~\eqref{mass},
\begin{equation}\label{eq:2}
    \begin{split}
        H&=\vec\sigma\cdot\Big(\frac{\partial}{\partial\vec x}+\vec a\Big)+m \sigma_3\\
    \end{split},
\end{equation}
where 
\begin{align}
\vec{a} = \Big(0, b x_1 + W \sin(K x_1) \Big).
\end{align}
To zeroth order in $W$, $H$ has the particle spectrum,
\[
E_n^{\left(0\right)} = 
\begin{cases}
  {\color{red}} \sqrt{2n |b|+ m^2},\quad n =1,2,\ldots, \cr
    \cr
  |m|,\quad n =0.
\end{cases}
\]
with the corresponding eigenfunctions,
\[
\psi_{n,p_2}(\vec{x}) = 
\begin{cases}
  \mathcal{N}e^{ip_2 x_2}
  \begin{pmatrix}
    -i\Phi_{n-1}\big(\frac{x_1+x_b}{l_b}\big)\\
    \frac{\sqrt{m^2+2n|b|}-m}{\sqrt{2n|b|}}\Phi_{n}\big(\frac{x_1+x_b}{l_b}\big)\\
    \end{pmatrix} & \text{for $n = 1, 2, \ldots$}, \\
    \\
  \mathcal{N}e^{ip_2 x_2}
  \begin{pmatrix}
    0\\
    \Phi_0\big(\frac{x_1+x_b}{l_b}\big)\\
    \end{pmatrix} & \text{for $n=0$},
\end{cases}
\]
where the normalization constant,
\begin{equation*}\label{eq:normalization}
    \mathcal{N}=\sqrt{\frac{n |b|}{l_bL_y(m^2+2n |b|-m\sqrt{m^2+2n |b|})}},
\end{equation*}
$k_2 \in {2 \pi \over L_2} \mathbb{Z}$ is the momentum along the $x_2$-direction ($L_2 \rightarrow \infty$), $x_b(p_2) \equiv x_b = p_2 l_b^2$, $l_b^{-1} = |b|$, and $\Phi_n(z) = {e^{-z^2/2} \over \sqrt{2^n n! \sqrt{\pi}}} H_n(z)$ for the $n$-th Hermite polynomial $H_n(z)$.
Thus, the states are labeled by $M = (n, p_2)$.
We are interested in how the periodic vector potential in Eq.~\eqref{backgroundvector} lifts the degeneracy of the flat Landau level spectrum and contributes to the velocity $v_i^M$.
(Finite dissipation has already been assumed in using a finite, non-zero scattering time $\tau$ in our calculation of the oscillatory component of $\rho_{xx}$.)

First order perturbation theory gives the energy level corrections,
\begin{equation}
    \begin{split}
        E_{n,p_2}^{(1)} = W\frac{\sqrt{2n}}{Kl_b}\Bigg[\sqrt{\frac{2n |b|}{m^2+2n |b|}}\Bigg]\cos(Kx_b)e^{-z/2}\Big[L_{n-1}(z)-L_n(z)\Big]\\
    \end{split},
\end{equation}
where $L_n(z)$ is the $n$th Laguerre polynomial, $z = K^2 l_b^2/2$, and terms suppressed as $L_1, L_2 \rightarrow \infty$ have been dropped.
Thus, to leading order, $v_1^{n, p_2} = 0$ and
\begin{equation}
     v_2^{n,p_2}=\frac{\partial E_{n,p_2}^{(1)}}{\partial p_2}=-W l_b\sqrt{2n}\Bigg[\sqrt{\frac{2n |b|}{m^2+2n |b|}}\Bigg]\sin(Kx_b)\mathrm{e}^{-z/2}\Big[L_{n-1}(z)-L_n(z)\Big].
\end{equation}
We substitute these $v_i^{n, p_2}$ into the Kubo formula \eqref{kubo} to find $\Delta \sigma_{yy}^\psi$.
To perform the integral over $p_2$, we approximate the Fermi-Dirac distribution function by substituting in the zeroth order energies $E_n^{(0)}$ (which are independent of $p_2$).
Thus, we obtain the periodic potential correction to the Dirac composite fermion conductivity:
\begin{equation}\label{eq:del_sig_1}
    \begin{split}
        \Delta\sigma_{yy}^{\psi}
        & \approx W^2\widetilde{\tau}\beta \sum_{n=0}^\infty\Bigg(\frac{2n |b|}{m^2+2n |b|}\Bigg)\frac{n\exp(\beta(E_{n}^{(0)}-\mu))}{\Big[1+\exp(\beta(E_{n}^{(0)}-\mu))\Big]^2}\mathrm{e}^{-z}\Big[L_{n-1}(z)-L_n(z)\Big]^2,
    \end{split}
\end{equation}
where $\tilde{\tau} \propto \tau$ has absorbed non-universal ${\cal O}(1)$ constants.

$\Delta \sigma_{yy}^\psi$ in Eq.~\eqref{eq:del_sig_1} exhibits both Shubnikov--de Haas (for large $|b|$) and Weiss oscillations (for smaller $|b|$).
We are interested in extracting an analytic expression that approximates Eq.~\eqref{eq:del_sig_1} at low temperatures near $\nu = 1/2$, following the earlier analysis in \cite{peetersvasilopoulosmagnetic}.
In the weak field limit, $|b|/\mu^2 \ll 1$, a large number of Landau levels are filled ($n\to\infty$).
Thus, we express the Laguerre polynomials $L_n$ as
\begin{equation}
    L_n\left(z\right)
    \xrightarrow[n \to \infty]{}
    \mathrm{e}^{z/2}\frac{\cos\left(2\sqrt{nz}-\frac{\pi}{4}\right)}{\left(\pi^2nz\right)^{1/4}} + {\cal O}({1 \over n^{3/4}}).
\end{equation}
Next, we take the continuum approximation for the summation over $n$ by substituting
\begin{equation*}
n \rightarrow {l_b^2 \over 2}\Big(E^2 - m^2 \Big), \quad  \sum_{n}\rightarrow l_b^2 \int\ EdE,
\end{equation*}
into Eq.~\eqref{eq:del_sig_1}:
\begin{align}
\label{approximatedconductivity}
\Delta\sigma_{yy}^{\psi} = {\cal C} \int_{- \infty}^{\infty} dE\ {\beta e^{\beta (E - \mu)} \over (1 + e^{\beta(E - \mu)})^2} \sin^2\Big(l_b^2 K \sqrt{E^2 - m^2} - {\pi \over 4} \Big),
\end{align}
where ${\cal C} = W^2 {\tilde \tau} l_b^2 K$ and we have approximated $2n |b|/(m^2 + 2 n |b|)$ by unity.
(The substitution for $n$ is motivated by the zeroth order expression for the energy of the Dirac composite fermion Landau levels.)
Anticipating that at sufficiently low temperatures the integrand in Eq.~\eqref{approximatedconductivity} is dominated by ``energies" $E$ near the Fermi energy $\mu$, we write:
\begin{align}
E = \mu + s T
\end{align}
so that Eq.~\eqref{approximatedconductivity} becomes for $|s| T \ll \mu = \sqrt{B}$:
\begin{align}
\label{thirdform}
\Delta\sigma_{yy}^{\psi} = {\cal C} \int_{- \infty}^{\infty} ds\ {e^{s} \over (1 + e^{s})^2} \sin^2\Big(l_b^2 K \sqrt{B - m^2} + {s T l_b^2 K \over \sqrt{1 - {m^2 \over B}}} - {\pi \over 4} \Big).
\end{align}
Performing the integral over $s$, we find the Weiss oscillations (see Eq.~\eqref{dictionaryoscillation}):
\begin{align}
\label{finiteTresult}
\Delta \rho_{xx} \propto 1 - {T/T_D \over \sinh(T/T_D)} \Big[1 - 2 \sin^2\Big({2 \pi l_b^2\sqrt{B - m^2} \over d} - {\pi \over 4}\Big)\Big],
\end{align}
where 
\begin{align}
T_D^{-1} = {4 \pi^2 l_b^2 \over d} {1 \over \sqrt{1 - {m^2 \over B}}},
\end{align}
we have substituted $K = 2\pi/d$, $l_b^2 = |b|^{-1}$, and the proportionality constant is controlled by the longitudinal resistivity at $\nu = 1/2$.

\begin{figure}
 \center
\includegraphics[scale=0.47]{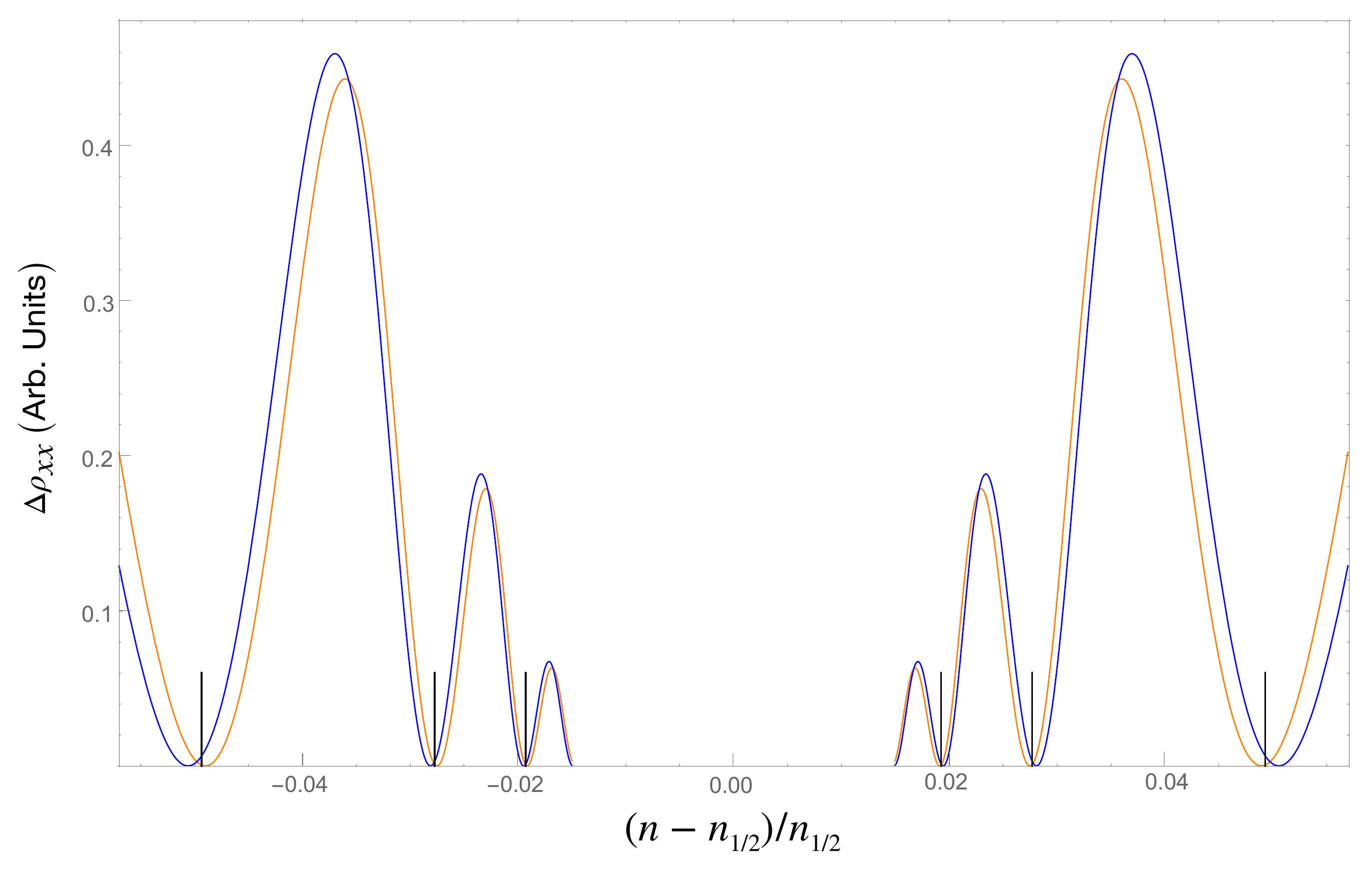}
\caption{Weiss oscillations of the Dirac composite fermion theory at fixed magnetic field $B$ and varying electron density $n_e$ about half-filling $n_{1/2} = B_{1/2}/4\pi$ ($\ell_{B_{1/2}}/d = 0.03$ and $k_B T = 0.3\sqrt{2 B_{1/2}}$).
The blue curve corresponds to Dirac composite fermion mean-field theory \cite{PhysRevB.95.235424}. 
The orange curve includes the effects of a Dirac composite fermion mass $m \propto |B - 4\pi n_e|^{1/3} B^{1/6}$ induced by gauge fluctuations.
Vertical lines correspond to the observed oscillation minima \cite{Kamburov2014}.}
\label{weissoscillationsfixedB}
 \end{figure}

Eq.~\eqref{finiteTresult} constitutes the primary result of this section.
The minima of $\Delta \rho_{xx}$ occur when
\begin{align}
{1 \over |b|} = {d \over 2 \sqrt{B - m^2}}\Big(p + {1 \over 4}\Big), p = 1, 2, 3, \ldots,
\end{align}
where $m$ is given in Eq.~\eqref{mass}.
For either fixed electron density $n_e$ or fixed external field $B$, the locations of the oscillation minima for a given $p$ (either $B(p)$ or $n_e(p)$) are shifted inwards towards $\nu = 1/2$. 
This is shown in Fig.~\ref{weissoscillations} for fixed $n_e$ and in Fig.~\ref{weissoscillationsfixedB} for fixed $B$.
The magnitude of this shift is symmetric for fixed $B$, but asymmetric for fixed $n_e$, given the form of the mass in Eq.~\eqref{mass}.
Mass dependence also appears in the temperature-dependent prefactor ${T/T_D \over \sinh(T/T_D)}$.
In principle, this mass dependence could be extracted from the finite-temperature scaling of $\Delta \rho_{xx}$ at the oscillation extrema.

\section{Comparison to HLR mean-field theory at finite temperature}
\label{discussion}

\subsection{Shubnikov--de Haas oscillations}

In \cite{Manoharan1994}, Shayegan et al.~found the Shubnikov--de Haas (SdH) oscillations near half-filling to be well described over two orders of magnitude in temperature by the formula, 
\begin{align}
\label{normalsdh}
{\Delta \rho_{xx} \over \rho_0} \propto {\xi_{NR} \over \sinh(\xi_{NR})} \cos(2 \pi \nu - \pi),
\end{align}
where $\xi_{NR} = {2 \pi^2 T \over \omega_c}$, $\omega_c = |b|/m^\ast$, $m^\ast$ is an effective mass, $\nu$ is the electron filling fraction, and $\rho_0$ is the longitudinal resistivity at half-filling (measured at the lowest accessible temperature).
(Note that these experiments were performed without any background periodic potential and so no Weiss oscillations were present.)
Recall that we are using units where $k_B = \hbar = e = c = 1$.
In particular, it was found that $m^\ast \propto \sqrt{B}$ for sufficiently large $|b| = |B - 4 \pi n_e|$ and that $m^\ast$ appeared to diverge as half-filling was approached.
Interpreted within the HLR composite fermion framework, $m^\ast$ corresponds to the composite fermion effective mass.
The $\sqrt{B}$ behavior of the composite fermion effective mass is consistent with the theoretical expectation \cite{halperinleeread, read1996cfs} that the composite fermion mass scale at $\nu=1/2$ is determined entirely by the characteristic energy of the Coulomb interaction.
(Away from $\nu=1/2$, scaling implies the effective mass can be a scaling function of $B$ and $n_e$.)

Applying previous treatments of SdH oscillations in graphene \cite{PhysRevB.71.125124, PhysRevB.78.245406} to the Dirac composite fermion theory, the temperature dependence of the SdH oscillations is controlled by
\begin{align}
{\Delta \rho_{xx} \over \rho_0} \propto {\xi_{D} \over \sinh(\xi_D)},
\end{align}
where $\xi_D = {2 \pi^2 T \sqrt{B} \over |b|}$.
Thus, $\xi_{NR} \propto \xi_D$ if $m^\ast \propto \sqrt{B}$. 
Consequently, the Dirac composite fermion theory is consistent with the observed temperature scaling with $\sqrt{B}$.
We cannot account for the divergence at small $|b|$ attributed to $m^\ast$ in our treatment.

\subsection{Weiss oscillations}

In \cite{PhysRevB.95.235424}, it was shown that the locations of the Weiss oscillation minima obtained from Dirac and HLR composite fermion mean-field theories coincide to $0.002\%$.
This result provides evidence that the two composite fermion theories may belong to the same universality class.
However, the (possible) equivalence of the two theories only occurs at long distances and so the finite-temperature behavior of the two theories will generally differ.

In HLR {\it mean-field theory}, the temperature dependence of the Weiss oscillations enters in the factor \cite{peetersvasilopoulosmagnetic},
\begin{align}
\Delta \rho_{xx} \propto {T/T_{NR} \over \sinh(T/T_{NR})},
\end{align}
where the characteristic temperature scale,
\begin{align}
T^{-1}_{NR} = {4 \pi^2 l_b^2 \over d} {m^\ast \over \sqrt{4 \pi n_e}}. 
\end{align}
Assuming the effective mass $m^\ast \propto \sqrt{B}$, the characteristic temperatures $T_D$ and $T_{NR}$ generally have very different behaviors as functions of $B$ and $n_e$.
It would be interesting to study the effects of fluctuations in HLR theory, along the lines of the study presented here, and compare with our result in Eq.~\eqref{finiteTresult}.

\section{Conclusion}
\label{conclusion}

In this paper, we studied theoretically commensurability oscillations about $\nu=1/2$ that are produced by a one-dimensional scalar potential using the Dirac composite fermion theory. 
Through an approximate large $N$ analysis of the Schwinger--Dyson equations, we considered how corrections to Dirac composite fermion mean-field theory affect the behavior of the predicted oscillations. 
We focused on corrections arising from the exchange of an emergent gauge field whose low-energy kinematics satisfy $|\vec{q}\,| \leq |q_0|$.
In addition, we only considered screened electron-electron interactions.
Remarkably within this restricted parameter regime, we found a self-consistent solution to the Schwinger--Dyson equations in which a Chern-Simons term for the gauge field and mass for the Dirac composite fermion are dynamically generated.
The Dirac mass resulted in a correction to the locations of the commensurability oscillation minima which improved comparison with experiment. 

There are a variety of directions for future exploration. 
It would be interesting to consider the effects of the exchange of emergent gauge fields with $|q_0| <  |\vec{q}\,|$.
In this regime, Landau damping of the ``magnetic" component of the gauge field propagator is expected to result in IR dominant Dirac composite fermion self-energy corrections \cite{SSLee2009OrderofLimits, MetlitskiSachdev2010Part1, Mross2010}.
In particular, it would be interesting to understand this regime when a dynamically-generated Chern-Simons term for the gauge field is present.
These studies are expected to be highly sensitive to the nature of the electron-electron interactions. 
At $\nu=1/2$ when the effective magnetic field vanishes, single-particle properties depend upon whether this interaction is short or long ranged \cite{KimFurusakiWenLee1994}.
It is important to understand the interplay of this physics with a non-zero effective magnetic field that is generated away from $\nu=1/2$ and its potential observable effects.

The corrections to the predicted commensurability oscillations relied on a solution to the Schwinger--Dyson equations, obtained in a large $N$ flavor approximation, that was extrapolated to $N=1$.
The study of higher-order in $1/N$ effects may provide additional insight into the validity of this extrapolation.
Alternatively, study of the 't Hooft large $N$ limit of the Dirac composite fermion theory dual conjectured in \cite{PhysRevB.99.125135} may complement our analysis.

Recent works \cite{2017PhRvX...7c1029W, 2018arXiv180307767K, PhysRevB.98.115105} have shown that PH symmetry at $\nu=1/2$ and reflection symmetry about $\nu=1/2$ rely on precisely correlated electric and magnetic perturbations. 
(This correlation is implemented by the Chern-Simons gauge field in the HLR theory.)
Specifically, a periodic scalar potential $V({\bf x})$ generates a periodic magnetic flux $b({\bf x})$ via 
\begin{align}
\label{slaving}
b({\bf x}) = - 2 m^\ast V({\bf x}).
\end{align}
How might fluctuations about HLR mean-field theory affect Eq.~\eqref{slaving} and potentially modify its predicted commensurability oscillations and other observables?

\section*{Acknowledgments}

We thank Hamed Asasi, Sudip Chakravarty, Pak Kau Lim, Leonid Pryadko, Srinivas Raghu, Nicholas Rombes, and Mansour Shayegan for useful conversations and correspondence.
M.M. is supported by the Department of Energy Office of Basic Energy Sciences contract DE-SC0020007.
M.M. and A.M. are supported by the UCR Academic Senate and the Hellman Foundation.
This work was performed in part at Aspen Center for Physics, which is supported by National Science Foundation grant PHY-1607611.

\appendix

\section{Integrals}
\label{integralappendix}

In this appendix, we give details for the calculations of the gauge and fermion self-energy integrals quoted in the main text.

\subsection{Gauge field self-energy}
\label{gaugefieldselfenergyappendix}

We begin with the gauge field self-energy given in \S\ref{gaugeselfenergy}.
We are interested in computing the PH odd component of the gauge field self-energy $\Pi_{\rm odd}$:
\begin{align}
\label{gaugeinvarianceapp}
\Pi^{\alpha \beta}(q) = \Pi^{\alpha \beta}_{\rm even}(q) + i \epsilon^{\alpha \beta \tau} q_\tau \Pi_{\rm odd}(q).
\end{align}
To leading order in $b$, we substitute $G(p) = G^{(0)}(p)$ from Eq.~\eqref{exactzero} with $\Sigma_\alpha = 0$ for $\alpha \in \{0, 1, 2\}$ into Eq.~\eqref{SDgauge}:
\begin{align}
i \epsilon^{\alpha \beta \tau} q_\tau \Pi_{\rm odd}(q) & = N \Big\{ \int {d^3 p \over (2 \pi)^3} {\rm tr}\Big[\gamma^\alpha G^{(0)}(p) \gamma^\beta G^{(0)}(p + q) \Big]\Big\}_{\rm odd} \cr
& = - N \Big\{ \int {d^3 p \over (2\pi)^3} {\rm tr}\Big[ \gamma^\alpha {i (\gamma^\sigma (p_\sigma + \mu_0 \delta_{\sigma, 0}) + \Sigma_m) \over (p_0 + \mu_0)^2 - p_i^2 - \Sigma_m^2} \gamma^\beta \cr
& \times {i (\gamma^\tau (p_\tau + q_\tau + \mu_0 \delta_{\tau, 0}) + \Sigma_m) \over (p_0 + q_0 + \mu_0)^2 - (p_i + q_i)^2 - \Sigma_m^2} \Big] \Big\}_{\rm odd}.
\end{align}
We have suppressed the $i \epsilon_{p_0}$ factor in Eq.~\eqref{exactzero} that defines the Feynman contour for the Minkowski-signature $p_0$ integration because we will will evaluate the above integral in Euclidean signature.
In subsequent sections of this appendix, we will likewise suppress the $i \epsilon_{p_0}$ factor for the same reason without further comment.
Recall that the factor of $N$ arises from the fermion loop over $N$ flavors of Dirac composite fermions and that $\mu_0 > 0$.

To leading order in the derivative expansion, i.e., $\Pi_{\rm odd}(q=0)$, the expression for $\Pi_{\rm odd}(0)$ simplifies to
\begin{align}
\Pi_{\rm odd}(0) = - 2 i N \Sigma_m \int {d^3 p \over (2\pi)^3} {1 \over \left(\left(p_0 + \mu_0\right)^2 - p_i^2 - \Sigma_m^2 \right)^2}.
\end{align}
Here, we have used the trace identities,
\begin{align}
\label{traceidentities}
{\rm tr}\Big[\gamma^\alpha \gamma^\beta \Big] & = 2 \eta^{\alpha \beta},\cr
{\rm tr}\Big[\gamma^\alpha \gamma^\beta \gamma^\tau \Big] & = - 2 i \epsilon^{\alpha \beta \tau}.
\end{align}
To compute this integral, we first Wick rotate, $p_0 \mapsto i (p_E)_3$ and $d^3 p \mapsto i d^3 p_E$, and then sequentially integrate over $(p_E)_3$ and the spatial momenta $(p_E)_i$ ($i = 1,2$) to find: 
\begin{align}
\label{CSlevelcalculation}
\Pi_{\rm odd}(0) & = 2 N \Sigma_m \int {d^3 p_E \over (2\pi)^3} {1 \over (i (p_E)_3 + \mu_0)^2 - (p_E)_i^2 - \Sigma_m^2)^2} \cr
& = 2 N \Sigma_m \int {d^3 p_E \over (2\pi)^3} {1 \over ((p_E)_3 - \omega_+)^2 ((p_E)_3 - \omega_-)^2} \cr
& = {N \Sigma_m \over 2} \int {d^2 p_E \over (2\pi)^2} {\Theta(|\Sigma_m| - \mu_0) + \Theta(\mu_0 - |\Sigma_m|) \Theta(|(p_E)_i| - \sqrt{\mu_0^2 - \Sigma_m^2}) \over (|(p_E)_i|^2 + \Sigma_m^2)^{3/2}} \cr
& = {N \over 4 \pi} \Big(\Theta(|\Sigma_m| - \mu_0) {\Sigma_m \over |\Sigma_m|} + \Theta(\mu_0 - |\Sigma_m|) {\Sigma_m \over \mu_0} \Big),
\end{align}
where the step function $ \Theta(|(p_E)_i| - \sqrt{\mu_0^2 - \Sigma_m^2})$ in the third line ensures the double poles $\omega_{\pm} = i (\mu_0 \pm \sqrt{(p_E)_i^2 + \Sigma_m^2})$ occur on opposite sides of the real $(p_E)_3$ axis.
Eq.~\eqref{CSlevelcalculation} implies that, for $\mu_0 > |\Sigma_m| > 0$, the gauge field obtains a correction to its propagator that corresponds to an effective Chern-Simons term with level,
\begin{align}
\label{CSlevelappendix}
k = {N \over 2} {\Sigma_m \over \mu_0}.
\end{align}

\subsection{Fermion self-energy}
\label{fermionselfenergyappendix}

Next, we calculate the fermion self-energies $\Sigma_m$ and $\Sigma_0$ quoted in \S\ref{fermionselfenergy}.

We begin with $\Sigma_m$.
Taking the trace of both sides of Eq.~\eqref{SDfermion} and setting $\delta q_\alpha = 0$, we find:
\begin{align}
\label{massselfenergystart1}
i \Sigma_m(q_{\rm FS}) & = i {\cal M}^{(0)}(q_{\rm FS}) + i {\cal M}^{(1)}(q_{\rm FS}),
\end{align}
where 
\begin{align}
i {\cal M}^{(0)}(q_{\rm FS}) & = {1 \over 2} \int {d^3 p \over (2\pi)^3} {\rm tr} \Big[ \gamma^\alpha G^{(0)}(p + q_{\rm FS}) \gamma^\beta \Big({2 \pi \over k} {\epsilon_{\alpha \beta \sigma} p^\sigma \over p^2} \Big) \Big], \\
i {\cal M}^{(1)}(q_{\rm FS}) & = {1 \over 2} \int {d^3 p \over (2\pi)^3} {\rm tr} \Big[ \gamma^\alpha G^{(1)}(p + q_{\rm FS}) \gamma^\beta \Big({2 \pi \over k} {\epsilon_{\alpha \beta \sigma} p^\sigma \over p^2} \Big) \Big],
\end{align}
$G^{(0}(p)$ and $G^{(1)}(p)$ are given in Eqs.~\eqref{exactzero} and \eqref{exactone}, $k$ is given in Eq.~\eqref{CSlevelappendix}, and $q_{\rm FS} = (0, \mu_0 \hat{n})$ for the unit vector $\hat{n}$ (e.g., $\hat{n} = (\cos(\varphi), \sin(\varphi))$ where $\varphi$ parameterizes a point on the Fermi surface) normal to the (assumed) spherical Fermi surface.
As before, we set $\Sigma_\alpha = 0$ for $\alpha \in \{0,1, 2\}$ and only retain $\Sigma_m$ when using $G^{(0)}(p)$ and $G^{(1)}(p)$ to evaluate ${\cal M}^{(0)}$ and ${\cal M}^{(1)}$, as well as $\Sigma_0$ below.
It is convenient to define $Q = (\mu_0, \mu_0 \hat{n})$ so that
\begin{align}
i {\cal M}^{(0)}(q_{\rm FS}) & = {1 \over 2} \int {d^3 p \over (2\pi)^3} {\rm tr} \Big[ \gamma^\alpha \Big({i (\gamma^\sigma (p + Q)_\sigma + \Sigma_m) \over (p + Q)^2 - \Sigma_m^2} \Big) \gamma^\beta \Big({2 \pi \over k} {\epsilon_{\alpha \beta \tau} p^\tau \over p^2} \Big) \Big], \\
i {\cal M}^{(1)}(q_{\rm FS}) & = {1 \over 2} \int {d^3 p \over (2\pi)^3} {\rm tr} \Big[ \gamma^\alpha \Big({i b ( \mathbb{I}(p + Q)_0 + \gamma^0 \Sigma_m )\over ((p + Q)^2 - \Sigma_m^2)^2} \Big) \gamma^\beta \Big({2 \pi \over k} {\epsilon_{\alpha \beta \tau} p^\tau \over p^2} \Big) \Big].
\end{align}

We first consider ${\cal M}^{(0)} = {\cal M}^{(0)}(q_{\rm FS})$.
Using the trace identities in Eq.~\eqref{traceidentities}, we find
\begin{align}
i {\cal M}^{(0)} & = {\pi i \over k} \int {d^3 p \over (2\pi)^3} {\rm tr} \Big[ \gamma^\alpha \Big({(\gamma^\sigma (p + Q)_\sigma + \Sigma_m) \over (p + Q)^2 - \Sigma_m^2} \Big) \gamma^\beta \Big({\epsilon_{\alpha \beta \tau} p^\tau \over p^2} \Big) \Big] \cr
& = - {4 \pi \over k} \int {d^3 p \over (2\pi)^3} {(p + Q)_\sigma p^\sigma \over ((p + Q)^2 - \Sigma_m^2) p^2}.
\end{align}
Next, we combine denominators using the Feynman parameter $x$ and then shift the integration by defining $\ell_\alpha = p_\alpha + Q_\alpha x$:
\begin{align}
i {\cal M}^{(0)} & = - {4 \pi \over k} \int {d^3 p \over (2\pi)^3} \int_0^1 dx {(p + Q)_\sigma p^\sigma \over (p^2 + 2 p \cdot Q x + Q^2 x - \Sigma_m^2 x)^2} \cr
& = - {4 \pi \over k} \int {d^3 \ell \over (2\pi)^3} \int_0^1 dx {\ell^2 + \ell \cdot Q (1 - 2 x) - x(1-x) Q^2 \over (\ell^2 + Q^2 x (1 - x) - \Sigma_m^2 x)^2} \cr
& = - {4 \pi \over k} \int {d^3 \ell \over (2\pi)^3} \int_0^1 dx {\ell^2 \over (\ell^2 - \Sigma_m^2 x)^2},
\end{align}
where we evaluated $Q^2 = 0$ and dropped the linear in $\ell$ term in the third line since it vanishes upon integration over $\ell$.
Next, we Wick rotate by taking $\ell_0 \mapsto i (\ell_E)_3$, $\ell^2 \mapsto - \ell^2_E$, and $d^3 \ell \mapsto i d^3 \ell_E$, integrate over $\ell_E$ via dimensional regularization, and finally integrate over $x$:
\begin{align}
i {\cal M}^{(0)} & = {4 \pi i \over k} \int {d^3 \ell_E \over (2\pi)^3} \int_0^1 dx {\ell_E^2 \over (\ell_E^2 + \Sigma_m^2 x)^2} \cr
& = - {12 \pi^{3/2} i |\Sigma_m| \over k (4 \pi)^{3/2}} \int_0^1 dx \ x^{1/2} \cr
& = - i {|\Sigma_m| \over k} \cr
& = - i {2 \mu_0 {\rm sign}(\Sigma_m) \over N},
\end{align}
where we substituted in the Chern-Simons level given in Eq.~\eqref{CSlevelappendix} in the final line.

Next, consider ${\cal M}^{(1)} = {\cal M}^{(1)}(q_{\rm FS})$. 
Using the trace identities in Eq.~\eqref{traceidentities}, we find
\begin{align}
i {\cal M}^{(1)} & = - {4 \pi b \Sigma_m\over k} \int {d^3 p \over (2\pi)^3} {p_0 \over ((p + Q)^2 - \Sigma_m^2)^2 p^2} \cr
& = - {4 \pi b \Sigma_m\over k} I(\Sigma_m^2, Q).
\end{align}
With the help of the formal identity,
\begin{align}
I(\Sigma_m^2, Q) = - \partial_{\Sigma_m^2} J(\Sigma_m^2, Q) = - \partial_{\Sigma_m^2} \int {d^3 p \over (2\pi)^3} {p_0 \over ((p + Q)^2 - \Sigma_m^2) p^2},
\end{align}
we rewrite
\begin{align}
i {\cal M}^{(1)} & = {4 \pi b \Sigma_m \over k} \partial_{\Sigma_m^2} \int {d^3 p \over (2\pi)^3} {p_0 \over ((p + Q)^2 - \Sigma_m^2) p^2}.
\end{align}
This integral has the same basic form as the one we encountered in calculating ${\cal M}^{(0)}$ and so we will follow the same steps as before: combine denominators with the Feynman parameter $x$, shift the integration $\ell_\alpha = p_\alpha + Q_\alpha x$, and substitute in $Q_0 = \mu_0$ and $Q^2 = 0$:
\begin{align}
i {\cal M}^{(1)} & = - {4 \pi b \Sigma_m \mu_0 \over k} \partial_{\Sigma_m^2} \int {d^3 \ell \over (2\pi)^3} \int_0^1 dx {x \over (\ell^2 - \Sigma_m^2 x)^2}.
\end{align}
Next, we Wick rotate by taking $\ell_0 \mapsto i (\ell_E)_3$, integrate over $\ell_E$ via dimensional regularization, integrate over $x$, take the derivative with respect to $\Sigma_m^2$, and then evaluate $k = {N \over 2} {\Sigma_m \over \mu_0}$:
\begin{align}
i {\cal M}^{(1)} & = - i {4 \pi b \Sigma_m \mu_0 \over k} \partial_{\Sigma_m^2} \int {d^3 \ell_E \over (2\pi)^3} \int_0^1 dx {x \over (\ell_E^2 + \Sigma_m^2 x)^2} \cr
& = - i {b \Sigma_m \mu_0 \over k} \partial_{\Sigma_m^2} {1 \over (\Sigma_m^2)^{1/2}} \int_0^1 dx\ x^{1/2} \cr
& = i {2 \over 3} {b \mu_0^2 \over N |\Sigma_m|^3}.
\end{align}

Finally, we calculate $\Sigma_0(q_{\rm FS})$ and $\Sigma'_0(q_{\rm FS})$, which we obtain from evaluating the derivative with respect to $q_0$ of $\Sigma_0(P)$ at the Fermi surface:
\begin{align}
i \Sigma_0(P)  = {1 \over 2} \int {d^3 p \over (2\pi)^3} {\rm tr} \Big[\gamma^0 \gamma^\alpha G^{(0)}(p + P) \gamma^\beta \Big({2 \pi \over k} {\epsilon_{\alpha \beta \sigma} p^\sigma \over p^2} \Big) \Big],
\end{align}
where $P = (q_0 + \mu_0, \mu_0 \hat{n})$.
First, we note that 
\begin{align}
{\rm tr}[\gamma^0 \gamma^\alpha \gamma^\sigma \gamma^\beta] (p + P)_\sigma p^\tau \epsilon_{\alpha \beta \tau} & = 2 (\eta^{0 \alpha} \eta^{\sigma \beta} - \eta^{0 \sigma} \eta^{\alpha \beta} + \eta^{0 \beta} \eta^{\alpha \sigma}) (p + P)_\sigma p^\tau \epsilon_{\alpha \beta \tau} \cr
& = (p + P)^\beta p^\tau \epsilon_{0 \beta \tau} + (p + P)^\alpha p^\tau \epsilon_{\alpha 0 \tau} \cr
& = 0.
\end{align}
Therefore, only the term proportional to $\Sigma_m$ in the numerator of $G^{(0)}$ contributes.
Using the trace identities in Eq.~\eqref{traceidentities}, we find
\begin{align}
i \Sigma_0(P) & = {4 \pi \Sigma_m \over k} \int {d^3 p \over (2\pi)^3} {p^0 \over ((p + P)^2 - \Sigma_m^2) p^2}.
\end{align}
As above, we combine denominators, shift the integration variable $\ell_\alpha = p_\alpha + P_\alpha x$, and drop any linear in $\ell$ terms in the numerator:
\begin{align}
i \Sigma_0(P) & = - {4 \pi \Sigma_m (q_0 + \mu_0) \over k} \int {d^3 \ell \over (2\pi)^3} \int_0^1 dx {x \over (\ell^2 + x (1- x) P^2 - \Sigma_m^2 x)^2}.
\end{align}
We assume $\Sigma_m^2 > |P^2| \approx |2 \mu q_0|$.
Wick rotating $\ell_0 \mapsto i (\ell_E)_3$ and sequentially performing the $\ell_E$ and $x$ integrals, we find:
\begin{align}
i \Sigma_0(P) & = - i {4 \pi \Sigma_m (q_0 + \mu_0) \over k} \int {d^3 \ell_E \over (2\pi)^3} \int_0^1 dx {x \over (\ell_E^2 - x (1- x)(2 \mu_0 q_0) + \Sigma_m^2 x)^2} \cr
& =  - i {\Sigma_m (q_0 + \mu_0) \over 2 k} \int_0^1 dx {x \over (\Sigma_m^2 x - 2 \mu_0 q_0 x (1- x))^{1/2}} \cr
& = - i {\Sigma_m (q_0 + \mu_0) \over 3 k |\Sigma_m|} \Big(1 + {2 \mu_0 q_0 \over 5 |\Sigma_m|^2} + {\cal O}(q_0^2)\Big) \cr
& = - i {2 \mu_0 \over 3 N |\Sigma_m|} (q_0 + \mu_0) \Big(1 + {2 \mu_0 q_0 \over 5 |\Sigma_m|^2} + {\cal O}(q_0^2)\Big).
\end{align}
Taking the derivative of $\Sigma_0(P)$ with respect to $q_0$, evaluating at $q = (0, \mu_0 \hat{n})$, and retaining only the first term ($\mu_0 q_0 \ll |\Sigma_m|^2$), we obtain
\begin{align}
i \Sigma'_0(q_{\rm FS}) = - i {2 \mu_0 \over 3 N |\Sigma_m|}.
\end{align}

\bibliography{bigbib}{}

\providecommand{\href}[2]{#2}\begingroup\raggedright\begin{thebibliography}{10}

\bibitem{PhysRevLett.50.1219}
D.~Yoshioka, B.~I. Halperin, and P.~A. Lee, ``Ground state of two-dimensional
  electrons in strong magnetic fields and $\frac{1}{3}$ quantized hall
  effect,'' \href{http://dx.doi.org/10.1103/PhysRevLett.50.1219}{{\em Phys.
  Rev. Lett.} {\bfseries 50} (Apr, 1983) 1219--1222}.
  \url{https://link.aps.org/doi/10.1103/PhysRevLett.50.1219}.

\bibitem{girvin1984}
S.~M. Girvin, ``Particle-hole symmetry in the anomalous quantum hall effect,''
  \href{http://dx.doi.org/10.1103/PhysRevB.29.6012}{{\em Phys. Rev. B}
  {\bfseries 29} (May, 1984) 6012--6014}.

\bibitem{Shahar1995}
D.~Shahar, D.~C. Tsui, M.~Shayegan, R.~N. Bhatt, and J.~E. Cunningham,
  ``{Universal Conductivity at the Quantum Hall Liquid to Insulator
  Transition},'' \href{http://dx.doi.org/10.1103/PhysRevLett.74.4511}{{\em
  Phys. Rev. Lett.} {\bfseries 74} (1995) 4511}.
  \url{http://link.aps.org/doi/10.1103/PhysRevLett.74.4511}.

\bibitem{Wong1996}
L.~W. Wong, H.~W. Jiang, and W.~J. Schaff, ``Universality and phase diagram
  around half-filled landau levels,''
  \href{http://dx.doi.org/10.1103/PhysRevB.54.R17323}{{\em Phys. Rev. B}
  {\bfseries 54} (1996) R17323}.
  \url{http://link.aps.org/doi/10.1103/PhysRevB.54.R17323}.

\bibitem{Pan2019}
W.~{Pan}, W.~{Kang}, M.~P. {Lilly}, J.~L. {Reno}, K.~W. {Baldwin}, K.~W.
  {West}, L.~N. {Pfeiffer}, and D.~C. {Tsui}, ``{Particle-Hole Symmetry and the
  Fractional Quantum Hall Effect in the Lowest Landau Level},'' {\em arXiv
  e-prints} (Feb, 2019) arXiv:1902.10262,
  \href{http://arxiv.org/abs/1902.10262}{{\ttfamily arXiv:1902.10262
  [cond-mat.mes-hall]}}.

\bibitem{rezayi2000}
E.~H. Rezayi and F.~D.~M. Haldane, ``{Incompressible Paired Hall State, Stripe
  Order, and the Composite Fermion Liquid Phase in Half-Filled Landau
  Levels},'' \href{http://dx.doi.org/10.1103/PhysRevLett.84.4685}{{\em Phys.
  Rev. Lett.} {\bfseries 84} (May, 2000) 4685}.

\bibitem{Geraedtsetal2015}
S.~D. {Geraedts}, M.~P. {Zaletel}, R.~S.~K. {Mong}, M.~A. {Metlitski},
  A.~{Vishwanath}, and O.~I. {Motrunich}, ``{The half-filled Landau level: The
  case for Dirac composite fermions},''
  \href{http://dx.doi.org/10.1126/science.aad4302}{{\em Science} {\bfseries
  352} (2016) 197}, \href{http://arxiv.org/abs/1508.04140}{{\ttfamily
  arXiv:1508.04140 [cond-mat.str-el]}}.

\bibitem{halperinleeread}
B.~I. Halperin, P.~A. Lee, and N.~Read, ``{Theory of the half-filled Landau
  level},'' \href{http://dx.doi.org/10.1103/PhysRevB.47.7312}{{\em Phys. Rev.
  B} {\bfseries 47} (Mar, 1993) 7312--7343}.
  \url{http://link.aps.org/doi/10.1103/PhysRevB.47.7312}.

\bibitem{Willett97}
R.~L. Willett, ``{Experimental evidence for composite fermions},'' {\em
  Semicond. Sci. Technol.} {\bfseries 12} (1997) 495.

\bibitem{Jainbook}
J.~K. Jain, {\em {Composite Fermions}}.
\newblock Cambridge University Press, 2007.

\bibitem{Fradkinbook}
E.~Fradkin, {\em {Field Theories of Condensed Matter Physics}}.
\newblock Cambridge University Press, 2013.

\bibitem{kivelson1997}
S.~A. Kivelson, D.-H. Lee, Y.~Krotov, and J.~Gan, ``{Composite-fermion Hall
  conductance at $\nu = 1/2$},''
  \href{http://dx.doi.org/10.1103/PhysRevB.55.15552}{{\em Phys. Rev. B}
  {\bfseries 55} (1997) 15552}.
  \url{http://link.aps.org/doi/10.1103/PhysRevB.55.15552}.

\bibitem{BMF2015}
M.~Barkeshli, M.~Mulligan, and M.~P.~A. Fisher, ``{Particle-hole symmetry and
  the composite Fermi liquid},''
  \href{http://dx.doi.org/10.1103/PhysRevB.92.165125}{{\em Phys. Rev. B}
  {\bfseries 92} (2015) 165125}.
  \url{http://link.aps.org/doi/10.1103/PhysRevB.92.165125}.

\bibitem{Son2015}
D.~T. Son, ``{Is the Composite Fermion a Dirac Particle?},''
  \href{http://dx.doi.org/10.1103/PhysRevX.5.031027}{{\em Phys. Rev. X}
  {\bfseries 5} (2015) 031027}.

\bibitem{2018arXiv181005174S}
T.~{Senthil}, D.~{Thanh Son}, C.~{Wang}, and C.~{Xu}, ``{Duality between
  $(2+1)d$ Quantum Critical Points},'' {\em ArXiv e-prints} (Oct., 2018)
  arXiv:1810.05174, \href{http://arxiv.org/abs/1810.05174}{{\ttfamily
  arXiv:1810.05174 [cond-mat.str-el]}}.

\bibitem{2017PhRvX...7c1029W}
C.~{Wang}, N.~R. {Cooper}, B.~I. {Halperin}, and A.~{Stern}, ``{Particle-Hole
  Symmetry in the Fermion-Chern-Simons and Dirac Descriptions of a Half-Filled
  Landau Level},'' \href{http://dx.doi.org/10.1103/PhysRevX.7.031029}{{\em
  Physical Review X} {\bfseries 7} no.~3, (July, 2017) 031029},
  \href{http://arxiv.org/abs/1701.00007}{{\ttfamily arXiv:1701.00007
  [cond-mat.str-el]}}.

\bibitem{2018arXiv180307767K}
P.~{Kumar}, S.~{Raghu}, and M.~{Mulligan}, ``{Composite fermion Hall
  conductivity and the half-filled Landau level},'' {\em arXiv e-prints} (Mar.,
  2018) arXiv:1803.07767, \href{http://arxiv.org/abs/1803.07767}{{\ttfamily
  arXiv:1803.07767 [cond-mat.str-el]}}.

\bibitem{PhysRevB.98.115105}
P.~Kumar, M.~Mulligan, and S.~Raghu, ``Topological phase transition
  underpinning particle-hole symmetry in the halperin-lee-read theory,''
  \href{http://dx.doi.org/10.1103/PhysRevB.98.115105}{{\em Phys. Rev. B}
  {\bfseries 98} (Sep, 2018) 115105}.
  \url{https://link.aps.org/doi/10.1103/PhysRevB.98.115105}.

\bibitem{PhysRevLett.117.216403}
S.~Golkar, D.~X. Nguyen, M.~M. Roberts, and D.~T. Son, ``Higher-spin theory of
  the magnetorotons,''
  \href{http://dx.doi.org/10.1103/PhysRevLett.117.216403}{{\em Phys. Rev.
  Lett.} {\bfseries 117} (Nov, 2016) 216403}.
  \url{https://link.aps.org/doi/10.1103/PhysRevLett.117.216403}.

\bibitem{PhysRevB.95.235424}
A.~K.~C. Cheung, S.~Raghu, and M.~Mulligan, ``{Weiss oscillations and
  particle-hole symmetry at the half-filled Landau level},''
  \href{http://dx.doi.org/10.1103/PhysRevB.95.235424}{{\em Phys. Rev. B}
  {\bfseries 95} (Jun, 2017) 235424}.
  \url{https://link.aps.org/doi/10.1103/PhysRevB.95.235424}.

\bibitem{PhysRevB.99.205151}
P.~Kumar, M.~Mulligan, and S.~Raghu, ``Emergent reflection symmetry from
  nonrelativistic composite fermions,''
  \href{http://dx.doi.org/10.1103/PhysRevB.99.205151}{{\em Phys. Rev. B}
  {\bfseries 99} (May, 2019) 205151}.
  \url{https://link.aps.org/doi/10.1103/PhysRevB.99.205151}.

\bibitem{LevinSon2016}
M.~Levin and D.~T. Son, ``{Particle-hole symmetry and electromagnetic response
  of a half-filled Landau level},''
  \href{http://dx.doi.org/10.1103/PhysRevB.95.125120}{{\em Phys. Rev. B}
  {\bfseries 95} (Mar, 2017) 125120}.
  \url{https://link.aps.org/doi/10.1103/PhysRevB.95.125120}.

\bibitem{PhysRevB.94.245107}
C.~Wang and T.~Senthil, ``Composite fermi liquids in the lowest landau level,''
  \href{http://dx.doi.org/10.1103/PhysRevB.94.245107}{{\em Phys. Rev. B}
  {\bfseries 94} (Dec, 2016) 245107}.
  \url{https://link.aps.org/doi/10.1103/PhysRevB.94.245107}.

\bibitem{BalramRifmmodeCsabaJain2015}
A.~C. Balram, C.~T\ifmmode~\mbox{\H{o}}\else \H{o}\fi{}ke, and J.~K. Jain,
  ``{Luttinger Theorem for the Strongly Correlated Fermi Liquid of Composite
  Fermions},'' \href{http://dx.doi.org/10.1103/PhysRevLett.115.186805}{{\em
  Phys. Rev. Lett.} {\bfseries 115} (2015) 186805}.
  \url{http://link.aps.org/doi/10.1103/PhysRevLett.115.186805}.

\bibitem{Weissfirst}
D.~Weiss, K.~V. Klitzing, K.~Ploog, and G.~Weimann, ``Magnetoresistance
  oscillations in a two-dimensional electron gas induced by a submicrometer
  periodic potential,'' {\em EPL (Europhysics Letters)} {\bfseries 8} no.~2,
  (1989) 179. \url{http://stacks.iop.org/0295-5075/8/i=2/a=012}.

\bibitem{gerhardtsweissklitzing}
R.~R. Gerhardts, D.~Weiss, and K.~v. Klitzing, ``Novel magnetoresistance
  oscillations in a periodically modulated two-dimensional electron gas,''
  \href{http://dx.doi.org/10.1103/PhysRevLett.62.1173}{{\em Phys. Rev. Lett.}
  {\bfseries 62} (Mar, 1989) 1173--1176}.
  \url{http://link.aps.org/doi/10.1103/PhysRevLett.62.1173}.

\bibitem{winkler1989landau}
R.~Winkler, J.~Kotthaus, and K.~Ploog, ``Landau band conductivity in a
  two-dimensional electron system modulated by an artificial one-dimensional
  superlattice potential,'' {\em Physical review letters} {\bfseries 62}
  no.~10, (1989) 1177.

\bibitem{Weiss1990}
D.~Weiss, ``Magnetoquantum oscillations in a lateral superlattice,'' in {\em
  Electronic properties of multilayers and low-dimensional semiconductor
  structures}, pp.~133--150.
\newblock Springer, 1990.

\bibitem{peetersvasilopoulos1992scalar}
F.~M. Peeters and P.~Vasilopoulos, ``Electrical and thermal properties of a
  two-dimensional electron gas in a one-dimensional periodic potential,''
  \href{http://dx.doi.org/10.1103/PhysRevB.46.4667}{{\em Phys. Rev. B}
  {\bfseries 46} (Aug, 1992) 4667--4680}.
  \url{http://link.aps.org/doi/10.1103/PhysRevB.46.4667}.

\bibitem{zhanggerhardts}
C.~Zhang and R.~R. Gerhardts, ``Theory of magnetotransport in two-dimensional
  electron systems with unidirectional periodic modulation,''
  \href{http://dx.doi.org/10.1103/PhysRevB.41.12850}{{\em Phys. Rev. B}
  {\bfseries 41} (Jun, 1990) 12850--12861}.
  \url{http://link.aps.org/doi/10.1103/PhysRevB.41.12850}.

\bibitem{peetersvasilopoulosmagnetic}
F.~M. Peeters and P.~Vasilopoulos, ``Quantum transport of a two-dimensional
  electron gas in a spatially modulated magnetic field,''
  \href{http://dx.doi.org/10.1103/PhysRevB.47.1466}{{\em Phys. Rev. B}
  {\bfseries 47} (Jan, 1993) 1466--1473}.
  \url{http://link.aps.org/doi/10.1103/PhysRevB.47.1466}.

\bibitem{gerhardts1996}
R.~R. Gerhardts, ``Quasiclassical calculation of magnetoresistance oscillations
  of a two-dimensional electron gas in spatially periodic magnetic and
  electrostatic fields,''
  \href{http://dx.doi.org/10.1103/PhysRevB.53.11064}{{\em Phys. Rev. B}
  {\bfseries 53} (Apr, 1996) 11064--11075}.
  \url{http://link.aps.org/doi/10.1103/PhysRevB.53.11064}.

\bibitem{Kamburov2014}
D.~Kamburov, Y.~Liu, M.~A. Mueed, M.~Shayegan, L.~N. Pfeiffer, K.~W. West, and
  K.~W. Baldwin, ``{What Determines the Fermi Wave Vector of Composite
  Fermions?},'' \href{http://dx.doi.org/10.1103/PhysRevLett.113.196801}{{\em
  Phys. Rev. Lett.} {\bfseries 113} (2014) 196801}.
  \url{http://link.aps.org/doi/10.1103/PhysRevLett.113.196801}.

\bibitem{shayeganreview2019}
M.~Shayegan, ``Probing composite fermions near half-filled landau levels.''.

\bibitem{Itzykson:1980rh}
C.~Itzykson and J.~B. Zuber, {\em {Quantum Field Theory}}.
\newblock International Series In Pure and Applied Physics. McGraw-Hill, New
  York, 1980.

\bibitem{Miransky:2015ava}
V.~A. Miransky and I.~A. Shovkovy, ``{Quantum field theory in a magnetic field:
  From quantum chromodynamics to graphene and Dirac semimetals},''
  \href{http://dx.doi.org/10.1016/j.physrep.2015.02.003}{{\em Phys. Rept.}
  {\bfseries 576} (2015) 1--209},
\href{http://arxiv.org/abs/1503.00732}{{\ttfamily arXiv:1503.00732 [hep-ph]}}.

\bibitem{Gusynin:1995nb}
V.~P. Gusynin, V.~A. Miransky, and I.~A. Shovkovy, ``{Dimensional reduction and
  catalysis of dynamical symmetry breaking by a magnetic field},''
  \href{http://dx.doi.org/10.1016/0550-3213(96)00021-1}{{\em Nucl. Phys.}
  {\bfseries B462} (1996) 249--290},
\href{http://arxiv.org/abs/hep-ph/9509320}{{\ttfamily arXiv:hep-ph/9509320
  [hep-ph]}}.

\bibitem{MetlitskiVishwanath2016}
M.~A. {Metlitski} and A.~{Vishwanath}, ``{Particle-vortex duality of
  two-dimensional Dirac fermion from electric-magnetic duality of
  three-dimensional topological insulators},''
  \href{http://dx.doi.org/10.1103/PhysRevB.93.245151}{{\em Phys. Rev. B}
  {\bfseries 93} no.~24, (June, 2016) 245151},
  \href{http://arxiv.org/abs/1505.05142}{{\ttfamily arXiv:1505.05142
  [cond-mat.str-el]}}.

\bibitem{WangSenthilfirst2015}
C.~Wang and T.~Senthil, ``Dual dirac liquid on the surface of the electron
  topological insulator,''
  \href{http://dx.doi.org/10.1103/PhysRevX.5.041031}{{\em Phys. Rev. X}
  {\bfseries 5} (2015) 041031}.

\bibitem{KMTW2015}
S.~Kachru, M.~Mulligan, G.~Torroba, and H.~Wang, ``{Mirror symmetry and the
  half-filled Landau level},''
  \href{http://dx.doi.org/10.1103/PhysRevB.92.235105}{{\em Phys. Rev. B}
  {\bfseries 92} (2015) 235105}.
  \url{http://link.aps.org/doi/10.1103/PhysRevB.92.235105}.

\bibitem{MrossAliceaMotrunichexplicitderivation2016}
D.~F. Mross, J.~Alicea, and O.~I. Motrunich, ``{Explicit Derivation of Duality
  between a Free Dirac Cone and Quantum Electrodynamics in ($2+1$)
  Dimensions},'' \href{http://dx.doi.org/10.1103/PhysRevLett.117.016802}{{\em
  Phys. Rev. Lett.} {\bfseries 117} (Jun, 2016) 016802}.

\bibitem{MurthyShankar2016halfull}
G.~Murthy and R.~Shankar, ``$\ensuremath{\nu}=\frac{1}{2}$ landau level:
  Half-empty versus half-full,''
  \href{http://dx.doi.org/10.1103/PhysRevB.93.085405}{{\em Phys. Rev. B}
  {\bfseries 93} (2016) 085405}.
  \url{http://link.aps.org/doi/10.1103/PhysRevB.93.085405}.

\bibitem{Seiberg:2016gmd}
N.~{Seiberg}, T.~{Senthil}, C.~{Wang}, and E.~{Witten}, ``{A duality web in 2 +
  1 dimensions and condensed matter physics},''
  \href{http://dx.doi.org/10.1016/j.aop.2016.08.007}{{\em Annals of Physics}
  {\bfseries 374} (Nov., 2016) 395--433},
  \href{http://arxiv.org/abs/1606.01989}{{\ttfamily arXiv:1606.01989
  [hep-th]}}.

\bibitem{PhysRevX.6.031043}
A.~Karch and D.~Tong, ``Particle-vortex duality from 3d bosonization,''
  \href{http://dx.doi.org/10.1103/PhysRevX.6.031043}{{\em Phys. Rev. X}
  {\bfseries 6} (Sep, 2016) 031043}.
  \url{https://link.aps.org/doi/10.1103/PhysRevX.6.031043}.

\bibitem{2018arXiv181111367S}
J.~H. {Son}, J.-Y. {Chen}, and S.~{Raghu}, ``{Duality Web on a 3D Euclidean
  Lattice and Manifestation of Hidden Symmetries},'' {\em arXiv e-prints}
  (Nov., 2018) arXiv:1811.11367,
  \href{http://arxiv.org/abs/1811.11367}{{\ttfamily arXiv:1811.11367
  [hep-th]}}.

\bibitem{Schwinger:1951nm}
J.~S. Schwinger, ``{On gauge invariance and vacuum polarization},''
  \href{http://dx.doi.org/10.1103/PhysRev.82.664}{{\em Phys. Rev.} {\bfseries
  82} (1951) 664--679}.
[,116(1951)].

\bibitem{Ritus:1978cj}
V.~I. Ritus, ``{Method of Eigenfunctions and Mass Operator in Quantum
  Electrodynamics of a Constant Field},'' {\em Sov. Phys. JETP} {\bfseries 48}
  (1978) 788.
[Zh. Eksp. Teor. Fiz.75,1560(1978)].

\bibitem{Gorbar:2013upa}
E.~V. Gorbar, V.~A. Miransky, I.~A. Shovkovy, and X.~Wang, ``{Radiative
  corrections to chiral separation effect in QED},''
  \href{http://dx.doi.org/10.1103/PhysRevD.88.025025}{{\em Phys. Rev.}
  {\bfseries D88} no.~2, (2013) 025025},
\href{http://arxiv.org/abs/1304.4606}{{\ttfamily arXiv:1304.4606 [hep-ph]}}.

\bibitem{watson2014quark}
P.~Watson and H.~Reinhardt, ``Quark gap equation in an external magnetic
  field,'' {\em Physical Review D} {\bfseries 89} no.~4, (2014) 045008.

\bibitem{2015EPJC...75..167K}
V.~R. {Khalilov} and I.~V. {Mamsurov}, ``{Polarization operator in the 2+1
  dimensional quantum electrodynamics with a nonzero fermion density in a
  constant uniform magnetic field},''
  \href{http://dx.doi.org/10.1140/epjc/s10052-015-3389-6}{{\em European
  Physical Journal C} {\bfseries 75} (Apr., 2015) 167},
  \href{http://arxiv.org/abs/1502.05355}{{\ttfamily arXiv:1502.05355
  [hep-th]}}.

\bibitem{2018arXiv180802140R}
N.~{Rombes} and S.~{Chakravarty}, ``{Specific heat and pairing of Dirac
  composite fermions in the half-filled Landau level},'' {\em arXiv e-prints}
  (Aug., 2018) arXiv:1808.02140,
  \href{http://arxiv.org/abs/1808.02140}{{\ttfamily arXiv:1808.02140
  [cond-mat.str-el]}}.

\bibitem{PhysRevD.29.2423}
R.~D. Pisarski, ``Chiral-symmetry breaking in three-dimensional
  electrodynamics,'' \href{http://dx.doi.org/10.1103/PhysRevD.29.2423}{{\em
  Phys. Rev. D} {\bfseries 29} (May, 1984) 2423--2426}.
  \url{https://link.aps.org/doi/10.1103/PhysRevD.29.2423}.

\bibitem{AppelquistNashWijewardhanaQED3}
T.~Appelquist, D.~Nash, and L.~C.~R. Wijewardhana, ``Critical behavior in
  (2+1)-dimensional qed,''
  \href{http://dx.doi.org/10.1103/PhysRevLett.60.2575}{{\em Phys. Rev. Lett.}
  {\bfseries 60} (Jun, 1988) 2575--2578}.
  \url{http://link.aps.org/doi/10.1103/PhysRevLett.60.2575}.

\bibitem{PhysRevLett.62.3024}
D.~Nash, ``Higher-order corrections in (2+1)-dimensional qed,''
  \href{http://dx.doi.org/10.1103/PhysRevLett.62.3024}{{\em Phys. Rev. Lett.}
  {\bfseries 62} (Jun, 1989) 3024--3026}.
  \url{https://link.aps.org/doi/10.1103/PhysRevLett.62.3024}.

\bibitem{PhysRevLett.93.206602}
F.~D.~M. Haldane, ``Berry curvature on the fermi surface: Anomalous hall effect
  as a topological fermi-liquid property,''
  \href{http://dx.doi.org/10.1103/PhysRevLett.93.206602}{{\em Phys. Rev. Lett.}
  {\bfseries 93} (Nov, 2004) 206602}.
  \url{https://link.aps.org/doi/10.1103/PhysRevLett.93.206602}.

\bibitem{Miransky:2001qs}
V.~Miransky, G.~Semenoff, I.~Shovkovy, and L.~Wijewardhana, ``{Color
  superconductivity and nondecoupling phenomena in (2+1)-dimensional QCD},''
  {\em Phys. Rev. D} {\bfseries 64} (2001) 025005,
\href{http://arxiv.org/abs/hep-ph/0103227}{{\ttfamily arXiv:hep-ph/0103227
  [hep-ph]}}.

\bibitem{PhysRevB.8.2649}
T.~Holstein, R.~E. Norton, and P.~Pincus, ``de haas-van alphen effect and the
  specific heat of an electron gas,''
  \href{http://dx.doi.org/10.1103/PhysRevB.8.2649}{{\em Phys. Rev. B}
  {\bfseries 8} (Sep, 1973) 2649--2656}.
  \url{https://link.aps.org/doi/10.1103/PhysRevB.8.2649}.

\bibitem{1999tald.conf..177D}
G.~V. {Dunne}, ``{Course 3: Aspects of Chern-Simons Theory},'' in {\em
  Topological Aspects of Low Dimensional Systems}, A.~{Comtet}, T.~{Jolicoeur},
  S.~{Ouvry}, and F.~{David}, eds., vol.~69, p.~177.
\newblock Jan, 1999.
\newblock \href{http://arxiv.org/abs/hep-th/9902115}{{\ttfamily
  arXiv:hep-th/9902115 [cond-mat]}}.

\bibitem{matulispeetersweiss2007}
A.~Matulis and F.~M. Peeters, ``Appearance of enhanced weiss oscillations in
  graphene: Theory,'' \href{http://dx.doi.org/10.1103/PhysRevB.75.125429}{{\em
  Phys. Rev. B} {\bfseries 75} (Mar, 2007) 125429}.
  \url{http://link.aps.org/doi/10.1103/PhysRevB.75.125429}.

\bibitem{tahirsabeehmagneticweiss2008}
M.~Tahir and K.~Sabeeh, ``Quantum transport of dirac electrons in graphene in
  the presence of a spatially modulated magnetic field,''
  \href{http://dx.doi.org/10.1103/PhysRevB.77.195421}{{\em Phys. Rev. B}
  {\bfseries 77} (May, 2008) 195421}.
  \url{http://link.aps.org/doi/10.1103/PhysRevB.77.195421}.

\bibitem{2016arXiv161004068B}
R.~{Burgos} and C.~{Lewenkopf}, ``{Weiss oscillations in graphene with a
  modulated height profile},'' {\em ArXiv e-prints} (Oct., 2016) ,
  \href{http://arxiv.org/abs/1610.04068}{{\ttfamily arXiv:1610.04068
  [cond-mat.mes-hall]}}.

\bibitem{charbonneauvVV1982}
M.~Charbonneau, K.~M. van Vliet, and P.~Vasilopoulos, ``Linear response theory
  revisited one-body response formulas and generalized boltzmann equations,''
  {\em J. Math. Phys.} {\bfseries 23} no.~318, (1982) .

\bibitem{Manoharan1994}
H.~C. Manoharan, M.~Shayegan, and S.~J. Klepper, ``{Signatures of a Novel Fermi
  Liquid in a Two-Dimensional Composite Particle Metal},'' {\em Phys. Rev.
  Lett.} {\bfseries 73} (1994) 3270.

\bibitem{read1996cfs}
N.~Read, ``Recent progress in the theory of composite fermions near
  even-denominator filling factors,''
  \href{http://dx.doi.org/http://dx.doi.org/10.1016/0039-6028(96)00318-4}{{\em
  Surface Science} {\bfseries 361} (1996) 7 -- 12}.
  \url{http://www.sciencedirect.com/science/article/pii/0039602896003184}.

\bibitem{PhysRevB.71.125124}
V.~P. Gusynin and S.~G. Sharapov, ``Magnetic oscillations in planar systems
  with the dirac-like spectrum of quasiparticle excitations. ii. transport
  properties,'' \href{http://dx.doi.org/10.1103/PhysRevB.71.125124}{{\em Phys.
  Rev. B} {\bfseries 71} (Mar, 2005) 125124}.
  \url{https://link.aps.org/doi/10.1103/PhysRevB.71.125124}.

\bibitem{PhysRevB.78.245406}
P.~Goswami, X.~Jia, and S.~Chakravarty, ``Quantum oscillations in graphene in
  the presence of disorder and interactions,''
  \href{http://dx.doi.org/10.1103/PhysRevB.78.245406}{{\em Phys. Rev. B}
  {\bfseries 78} (Dec, 2008) 245406}.
  \url{https://link.aps.org/doi/10.1103/PhysRevB.78.245406}.

\bibitem{SSLee2009OrderofLimits}
S.-S. Lee, ``{Low-energy effective theory of Fermi surface coupled with U(1)
  gauge field in $2+1$ dimensions},'' {\em Phys. Rev. B} {\bfseries 80} (2009)
  165102.

\bibitem{MetlitskiSachdev2010Part1}
M.~A. Metlitski and S.~Sachdev, ``{Quantum phase transitions of metals in two
  spatial dimensions. I. Ising-nematic order},'' {\em Phys. Rev. B} {\bfseries
  82} (2010) 075127.

\bibitem{Mross2010}
D.~F. Mross, J.~McGreevy, H.~Liu, and T.~Senthil, ``{Controlled expansion for
  certain non-Fermi-liquid metals},'' {\em Phys. Rev. B} {\bfseries 82} (2010)
  045121.

\bibitem{KimFurusakiWenLee1994}
Y.~B. Kim, A.~Furusaki, X.-G. Wen, and P.~A. Lee, ``Gauge-invariant response
  functions of fermions coupled to a gauge field,'' {\em Phys. Rev. B}
  {\bfseries 50} (1994) 17917.

\bibitem{PhysRevB.99.125135}
A.~Hui, E.-A. Kim, and M.~Mulligan, ``Non-abelian bosonization and modular
  transformation approach to superuniversality,''
  \href{http://dx.doi.org/10.1103/PhysRevB.99.125135}{{\em Phys. Rev. B}
  {\bfseries 99} (Mar, 2019) 125135}.
  \url{https://link.aps.org/doi/10.1103/PhysRevB.99.125135}.

\end{thebibliography}\endgroup
\bibliographystyle{utphys}

\end{document}